%% file: tcs.tex
\documentclass[number]{elsarticle}      

\usepackage{graphicx}
\usepackage{amsmath}
\usepackage{amsfonts}
\usepackage{amssymb}
\usepackage{amsthm}
\usepackage{mathabx}
\usepackage[mathscr]{euscript}
\usepackage{amscd}
\usepackage[utf8]{inputenc}
\usepackage{stmaryrd}
\usepackage{xcolor}
\usepackage[english]{babel}
\usepackage{natbib}

\newdefinition{defn}{Definition}
\newtheorem{prop}[defn]{Proposition}
\newtheorem{theo}[defn]{Theorem}
\newtheorem{lema}[defn]{Lemma}

\newtheorem{fact}[defn]{Fact}

\newtheorem{corolary}[defn]{Corollary}

\newtheorem{innerrefteo}{Theorem}

\newenvironment{refteo}[1]
  {\renewcommand\theinnerrefteo{#1}\innerrefteo}
  {\endinnerrefteo}

\newtheorem{innerrefprop}{Proposition}
\newenvironment{refprop}[1]
  {\renewcommand\theinnerrefprop{#1}\innerrefprop}
  {\endinnerrefprop}
  
\newtheorem{innerreffact}{Fact}
\newenvironment{reffact}[1]
  {\renewcommand\theinnerreffact{#1}\innerreffact}
  {\endinnerreffact}

\def\blfootnote{\xdef\@thefnmark{}\@footnotetext}

\newcommand{\citeonline}[1]{\citep{#1}}

\newcommand{\mDP}[1]{\mathtt{(DP{#1})}}
\newcommand{\DP}[1]{$\mDP{#1}$}
\newcommand{\CR}[1]{$\mathtt{(CR{#1})}$}
\newcommand{\GR}{$\mathtt{(GR)}$}
\newcommand{\FAITH}{$\mathtt{(Faith)}$}
\newcommand{\REC}[1]{$\mathtt{(Rec{#1})}$}
\newcommand{\LC}[1]{$\mathtt{(LC{#1})}$}

\renewcommand{\textcolor}[2]{#2}
\begin{document}

\title{Dynamic Preference Logic meets Iterated Belief Change: Representation Results and Postulates Characterization}


\author[ufba,cor1]{Marlo Souza}
\ead{msouza1@ufba.br}

\author[pucrs]{Renata Vieira}
\ead{renata.vieira@pucrs.br}

\author[ufrgs]{Álvaro Moreira}
\ead{alvaro.moreira@inf.ufrgs.br}

\cortext[cor1]{Corresponding author}

\address[ufba]{Institute of Mathematics and Statistics, Federal University of Bahia - UFBA\\ Av. Adhemar de Barros, S/N, Ondina - Salvador-BA, Brazil}

\address[pucrs]{Faculty of Informatics, Pontifical Catholic University of Rio Grande do Sul - PUCRS\\ 
Av. Ipiranga, 6681 - Porto Alegre-RS, Brazil}

\address[ufrgs]{Institute of Informatics, Federal University of Rio Grande do Sul- UFRGS\\ Av. Bento Gonçalves, 9500 - Porto Alegre-RS, Brazil} 


\date{Received: date / Accepted: date}

\begin{abstract}
AGM's belief revision is one of the main paradigms in the study of belief change operations. Recently, several logics for belief and information change have been proposed in the literature and used to encode belief change operations in rich and expressive semantic frameworks. While the connections of AGM-like operations and their encoding in dynamic doxastic logics have been studied before by the work of Segerberg, most works on the area of Dynamic Epistemic Logics (DEL) have not, to our knowledge, attempted to use those logics as tools to investigate mathematical properties of belief change operators. This work investigates how Dynamic Preference Logic, a logic in the DEL family, can be used to study properties of dynamic belief change operators, focusing on well-known postulates of iterated belief change.

\end{abstract}


\maketitle

\input{intro.tex}

\input{prelim.tex}

\input{dpl.tex}
\input{ibc.tex}

\section{Final Considerations}
\label{sec:fin}

This work has investigated representation results for well-known iterated belief change postulates using the framework of Dynamic Epistemic Logics. We have provided a set of axioms that encode these postulates within DPL and shown that, as a result of  the higher expressiveness of preference models in comparison to Grove's SOS, a characterisation of these postulates in DPL cannot be obtained. 

\textcolor{blue}{Further, we have provided a generalisation of the postulates using preference models (Definitions~\ref{def:dp1plus}, \ref{def:dp2plus}, \ref{prop:dp3plus}, \ref{def:dp4plus}, \ref{def:recplus}, \ref{def:cr3plus}, \ref{def:cr4plus}, and \ref{def:lcplus}) that coincide with the studied postulates in the  class of models similar to Grove's spheres (Facts~\ref{cor:CR1}, \ref{fact:dp1syn}, \ref{fact:dp2syn}, \ref{fact:dp3syn}, \ref{fact:dp4syn}, and \ref{fact:recsyn}), and shown that our encoding characterises these generalised postulates (Propositions~\ref{prop:dp1plus}, \ref{prop:CR2}, \ref{prop:CR3}, \ref{prop:CR4}, \ref{prop:Rec}, \ref{prop:cont3}, \ref{prop:BasicAxiom}, and \ref{prop:LC}). Finally, we have shown (Theorem~\ref{teo:correct}) that we can obtain a sound axiomatisation for DPL interpreted over a class of dynamic models by aggregating the axioms for each postulate satisfied by all dynamic operators in that class.} 

\textcolor{blue}{We  use the proposed axioms to obtain an axiomatisation of Preference Logic extended with dynamic modalities for Lexicographic Revision (Corollary~\ref{cor:axiomRU}) and Lexicographic Contraction (Proposition~ \ref{prop:LC}). We wish to point out that, while our work is concerned with single-agent belief changes, our results can be trivially extended to \textit{private changes} in the multi-agent case.}

\textcolor{blue}{Our results generalise \textcolor{blue}{previous semantic-based postulates} by demonstrating  are intrinsically linked to the structure of the model on which they are based, i.e., Grove's system of spheres for  classical propositional logic. To avoid terminological confusion, we point out that our results are concerned with belief change operations defined on well-founded preference models, which we simply call preference models in this work, and they may not be valid if one considers all preference models, since some belief change operators may not be well-defined on non-well-founded models \cite{girard2014belief,souzaphd}.}

A question that may arise from our work is whether the generalisation of Belief Change postulates from the semantic framework of Grove's systems of spheres to preference models is relevant from an epistemological point of view. 
We point out both that preference models have been extensively studied as models for non-monotonic reasoning and conditional logics~\cite{kraus1990nonmonotonic,kaci2005non,friedman1994complexity} and to model different mental attitudes, such as Preferences \cite{girard2008modal,lang}, Goals and Desires~\cite{souza2017dynamic}, and Obligations~\cite{liu:deontics}, and different notions of Belief \cite{BAL08}. 
 
\textcolor{blue}{It is of notice that preference models can be used to represent and reason about nested conditionals (or sets of conditionals), as those studied in conditional logics \cite{chellas1975basic,boutilier1994conditional}, which can represent introspective beliefs and their dynamics.  These conditionals cannot be expressed by means of Grove models, as these models are injective, in the sense used by Friedman and Halpern~\cite{friedman1994complexity}.}

\textcolor{blue}{
The reader may also argue whether Modal Logic is an adequate framework to study belief change. We point out that, firstly, modal analysis of belief and other mental attitudes abound in the literature and have proven to be a fruitful and powerful framework to study attitudes such as beliefs \cite{meyer2003modal,dignum,rao1991modeling}. Also, as De Rijke~\cite{de1994meeting} points out, Dynamic Logic is a standard tool to reason about states and transitions between states, two fundamental notions for any notion of change or dynamics, and can be analysed through  well-established tools from Modal Logic. As such, theories based on Dynamic Logic can be easily connected to well-established modal analysis of mental attitudes in the literature, providing a rich framework to study dynamic phenomena.}

While we have proved in this work that the axiomatisations provided are complete characterisations of the investigated postulates, by means of our proposed generalisations, we were not able to provide results about the completeness of the logic obtained by extending preference logic with a given dynamic operator characterised by a set of postulates. Inasmuch as the obtained axiomatisation is indeed complete for the example studied, i.e., Lexicographic Revision, it is still unclear if this will always be the case. Completeness results are not a trivial topic in Modal Logic and it is not clear that a general result guaranteeing completeness of the derived axiomatisation in regards to the extended logic can be obtained.

 As future work, we aim to investigate how these characterisations can be connected to syntactic representations of belief change operations. It is well known that dynamic operators can be encoded by means of transformations on priority graphs, a connection already studied by Liu~\cite{liu2011reasoning}, Souza et al.~\cite{souzakr} and others. Souza et al.~\cite{souza2019iterated,souza:dali19} have shown that some postulates - as originally defined in Iterated Belief Change - cannot be encoded by means of transformations on priority graphs unless we restrict our semantics to only consider some specific classes of preference models. We aim to investigate if the same holds for our generalisations and, if so, for which kind of models such a syntactic characterisation of iterated belief change postulates can be achieved.

\section*{Funding}
This study was financed in part by the Coordenação de Aperfeiçoamento de
Pessoal de Nível Superior - Brasil (CAPES) - Finance Code 001.


\bibliographystyle{elsarticle-num}
\bibliography{relatorio}   

\appendix
\input{proofs.tex}

\end{document}

%% file: intro.tex
\section{Introduction}
\label{intro}
\footnotetext{List of Abbreviations: DEL - Dynamic Epstemic Logic, DPL - Dynamic Preference Logic, SOS- System of Spheres, OCF - Ordinal Conditional Function, m.e. - modally equivalent, PDL - Propositional Dynamic Logic, DDL - Dynamic Doxastic Logic.}

Belief Change is the multidisciplinary area that studies how a doxastic agent comes to change her mind after acquiring new information. The most influential approach to Belief Change in the literature is the AGM paradigm  \cite{AGM}. 

\textcolor{blue}{
AGM defines belief change operations by structural constraints on how the beliefs of an agent should change. However, it has been argued in the literature, that belief change operations should be defined by means of changes in the agent's epistemic state - understood as more than the currently held (unconditional) beliefs but including also the agent's dispositions to believe 
\cite{SPO88,darwiche}. Aiming to extend the AGM framework to account for this idea, several works have established what became known as \textit{Iterated Belief Change}.}

While the rational constraints for changes in an agent's mental attitudes have been well-investigated in areas such as Epistemology and Logic, e.g., \cite{AGM,hansson1995changes}, 
the integration of belief change within the logics of beliefs and knowledge is a somewhat recent development. The first of such attempts was proposed by Segerberg~\cite{segerberg,SEG01} with his Dynamic Doxastic Logic (DDL).

\textcolor{blue}{
This shift from the extra-logical characterisation of belief change 
to its integration within a representation language has important expressiveness consequences. It also allows for the exploration of established results from Modal Logic to construct applications of the AGM belief change theory.} \textcolor{blue}{
The work of Segerberg, Lindström, and Rabinowicz on DDL \cite{segerberg,lindstrom:rabinowicz}, for example, show that the dynamics of introspective and higher-order beliefs poses problems to the AGM framework as they are not compatible with the AGM postulates. Their analysis of the difference between reasoning about the state in which beliefs are held (the point of evaluation) and reasoning about the state in which certain things are believed (the point of reference) 
corresponds directly to the analysis of the Ramsey Conditional Test by Baltag and Smets~\cite{BAL08} in the context of Dynamic Epistemic Logic.} \textcolor{blue}{
Such integration has also proven to be fruitful in the area of Interactive Epistemology. Reny~\cite{reny}, for example, shows that if the agent is allowed to revise her beliefs, backward induction is not supported by the common belief of rationality. Further, Board~\cite{board2004dynamic} shows that the framework of extensive-form epistemic games extended with belief change is a rich framework for unifying different competing notions in the area.
}

\textcolor{blue}{
As such, it is clear that incorporating belief change within epistemic and doxastic logics may provide useful theoretical insights on the phenomena related to belief dynamics, as the fact that the variety of doxastic attitudes may give rise to a variety of belief change operations satisfying their formal properties. However, the investigation of the mathematical properties satisfied by these belief change operators is rarely pursued by these works.}  
  
Recently, Girard \cite{girard2008modal,girard2014belief} proposes Dynamic Preference Logic (DPL) 
to study generalisations of belief revision \textit{a la} AGM \cite{girard2008modal,girard2014belief}. Some works have integrated well-known belief change operators within this logic \cite{girard2008modal,liu2011reasoning,van2007dynamic,souzakr} and have used them to study the dynamic behaviour of attitudes such as Preferences, Beliefs, and Intentions. Often, it is unclear whether these dynamic logics can be used to express the desirable properties of the belief change operators they study.
 
Souza et al.~\cite{souza:dali} have demonstrated that belief change postulates can be encoded within DPL, showing that any belief change operator satisfying these postulates induce the validity of some axioms in their corresponding dynamic logic. This result allows the use of DPL as a language to reason about classes of Belief Change operators. However, the authors were not able to completely characterise the postulates within DPL, i.e., to provide axioms in the dynamic logic that guarantee the satisfaction of such postulates by the corresponding belief change operator.

In this work, we study the relationship between the iterated belief change postulates satisfied by belief change operators and the axioms valid in DPL using these operators.  We provide generalisations of some known iterated belief change postulates to the context of relational preference models and provide representations of these postulates within DPL, which completely characterise them.

The representation results obtained in our work highlight  which relation-changing operators over a given class of preference models can be considered to provide the semantics of the logic, in a similar sense as to how frame properties can be defined by means of modal axioms for relational semantics.

\textcolor{blue}{We point out that in this work, we choose to employ our preference models as opposed to Grove's Systems of Spheres to study belief and its changes. The reason for this is that, firstly, preference-like models are more general structures and have been extensively studied as models for non-monotonic reasoning, conditional logics, and mental attitudes, as variant notions of belief ~\cite{kraus1990nonmonotonic,kaci2005non,lang,souza2017dynamic,liu:deontics,BAL08}. By employing these models to study belief change operators, we may generalise our results and the insights obtained with our study to change operations for other mental attitudes and diverse dynamic phenomena.} 

The following are the main contributions of our work: (i) Theorem~\ref{teo:correct}, \textcolor{blue}{which proves that we can combine the axiomatisations obtained for each postulate satisfied by a class of operators, resulting in a sound axiomatisation of the logic induced by this class of operators}, and (ii)  the representation results for the iterated belief change postulates provided in Propositions~\ref{prop:dp1plus}, \ref{prop:CR2}, \ref{prop:CR3}, \ref{prop:CR4}, \ref{prop:Rec}, \ref{prop:cont3}, \ref{prop:cont4}, and \ref{prop:LC}.


Theorem~\ref{teo:correct} also establishes a method for deriving axiomatizations of dynamic operators in DPL. Our method is different from the ones proposed by Van Benthem and Liu~\cite{pref}, and by  Aucher~\cite{aucher2010characterizing}, which are based on Propositional Dynamic Logic without iteration. It uses the extensively studied postulates from Belief Change to derive an axiomatisation of the logic with a given operation. As such, our method applies to a wider class of dynamic belief change operators, including those that cannot be encoded using Propositional Dynamic Logic programs, e.g., Ramachandram et al.'s \cite{ramachandran2012three} Lexicographic Contraction. 

This work is an extended version of the work presented at the 8th Brazilian Conference on Intelligent Systems - BRACIS 2019 \cite{souza:bracis19} and includes several developments that resulted from questions left open in our original work, as well as those from discussions during the event. As the main new developments not present in the BRACIS paper we list:
\begin{itemize}
\item the correction of the generalised postulates \DP{1a}, \DP{1b}, \DP{2a}, and \DP{2b} to account for the changes in the strict part $<$ of the preference relation;
\item the correction of the generalised postulate \REC{'} to  exclude unwanted links between $\neg\varphi$ and $\varphi$ worlds, consistent with Nayak et al.'s~\cite{nayak2003dynamic} original \REC{} postulate, in Definitions~\ref{def:dp1plus}, \ref{def:dp2plus} and \ref{def:recplus};
\item the extension of our analysis to postulates for iterated belief contraction, namely Chopra et al.'s \cite{chopra2008iterated} \CR{1} - \CR{4} and Ramachandran et al.'s \cite{ramachandran2012three} \LC{}, in Propositions~\ref{prop:dp1plus}, \ref{prop:CR2}, \ref{prop:cont3}, \ref{prop:cont4} and \ref{prop:LC};
\item an extended presentation and discussion of DPL and the results related to this logic which are important to our characterisation and discussions, in Section~\ref{sec:dyn};
\item a better formalisation of the semantics of DPL, in Definition~\ref{def:semantics}, which allows us to interpret the representation results in Section~\ref{sec:iter} as constraints on the class of models used to interpret the language;
\item generalisations of all results showing that the representation results can be used to define satisfaction of the analysed postulates within any class of preference models, in Section~\ref{sec:iter};
\item the proof that the original postulates cannot be represented within DPL, exemplified by Fact~\ref{fact:cr1} which can be restated for all the other postulates studied in this work;
\item a more general soundness result for the resulting axiomatisation, in Theorem~\ref{teo:correct}, showing a stronger relationship between proof systems and models for the logic.
\end{itemize}

This work is structured as follows:  in Section~\ref{sec:ibr}, we discuss AGM belief change and the results and postulates in the literature of dynamic and iterated belief change; in Section~\ref{sec:dyn}, we introduce DPL, a logic in the tradition of Dynamic Epistemic Logic recently applied to study belief change; in Section~\ref{sec:iter} we present our main results: we investigate the relationship between the iterated belief change postulates satisfied by belief change operators and the axioms valid in DPL, as well as providing generalisations of some postulates of belief change for the context of relational preference models. In Section~\ref{sec:rel}, we discuss the related work, and finally, in Section~\ref{sec:fin}, we present our final considerations. Proofs of the main results in this work can be found in the Appendix.

%% file: prelim.tex
\section{Preliminaries}\label{sec:ibr}

Let us consider a logic $\mathcal{L} = \langle L, Cn \rangle$ where $L$ is the logical language and $Cn:2^L\rightarrow 2^L$ is a consequence operator. In AGM's approach, the belief state of an agent is represented by a belief set, i.e. a consequentially closed set $B \subseteq L$, with $B=Cn(B)$  of $\mathcal{L}$-formulas. 

\textcolor{blue}{In this framework, a belief change operator is any operation $\star:2^L \times L \rightarrow 2^L$ that, given a belief set $B$ and some information $\varphi$, changes the belief set in some way.} AGM investigated three basic belief change operators: expansion, contraction, and revision. Belief expansion blindly integrates a new piece of information into the agent's beliefs. Belief contraction removes a currently believed sentence from the agent's set of beliefs, with minimal alterations. Finally, belief revision is the operation of integrating new information into an agent's beliefs while maintaining consistency. 

Among these basic operations, only expansion can be univocally defined.  The other two operations are defined by a set of rational constraints or postulates,  usually referred to as the AGM postulates or the G\"ardenfors postulates. These postulates define a class of suitable change operators representing different rational ways in which an agent can change her beliefs.
Let $B\subseteq L$ be a belief set and $\alpha,\beta \in L$ $\mathcal{L}$-formulas. For the revision operation ($\ast$), the authors introduce the following postulates:
%
\begin{enumerate}
   \item[]$\mathtt{(R1)}$ $B\ast\alpha = Cn(B\ast\alpha)$
   \item[]$\mathtt{(R2)}$ $\alpha\in B \ast \alpha$
   \item[]$\mathtt{(R3)}$ $B \ast \alpha \subseteq Cn(B \cup \{\alpha\})$
   \item[]$\mathtt{(R4)}$ If $\neg \alpha \not \in B$, then $B\ast\alpha = Cn(B \cup \{\alpha\})$
   \item[]$\mathtt{(R5)}$ $B\ast\alpha = Cn(\{\bot\})$ iff $\vdash \neg \alpha$
   \item[]$\mathtt{(R6)}$ If $\vdash \alpha \leftrightarrow \beta$ then $B\ast\alpha = B\ast\beta$
   \item[]$\mathtt{(R7)}$ $B \ast (\alpha \wedge \beta) \subseteq Cn((B\ast\alpha)\cup \{\beta\})$
   \item[]$\mathtt{(R8)}$ If $\neg \beta \not \in B\ast\alpha$, then $Cn((B\ast\alpha)\cup \{\beta\})\subseteq B\ast (\alpha \wedge \beta)$
\end{enumerate}

Belief expansion, contraction, and revision are interconnected by the pro\-per\-ties known as Levi and Harper identities \cite{AGM}.

While the AGM approach is independent of the supporting logic's syntax, it lacks a clear semantic interpretation for its operations. Grove~\cite{grove} provided one such interpretation using a possible-world semantics, based on Lewis' \cite{lewis} spheres. Grove's model for the operation of belief revision has clarified the meaning of belief change operations and has become a necessary tool for the development of new methods and operations in the area, such as the iterated belief operations we will discuss later.

A Grove system of spheres (SOS) is a pair $\mathcal{S} = \langle W, \leq\rangle$ where $W$ is a set of models for the logic $\mathcal{L}$ and $\leq\,  \subseteq W\times W$ satisfies the following conditions\footnote{Grove also requires the property of Universality that $W$ is a complete class of models for the logic $\mathcal{L}$.}: $\mathtt{(i)}$ $\leq$ is connected,  $\mathtt{(ii)}$ $\leq$ is transitive, and $\mathtt{(iii)}$ for any $S\subseteq W$, if $S \neq \emptyset$, then exists $x \in S$ minimal in $\leq$ in regards to $S$.

Given a system of spheres $\mathcal{S} = \langle W,\leq\rangle$, we say that the set of models of a formula $\varphi \in L$ is the set $\llbracket\varphi\rrbracket = \{w \in W~|~ w\vDash_\mathcal{L} \varphi\}$,  and for a set of models $\Gamma\subseteq W$ we say the set of minimal models of $\Gamma$ is: $$Min_\leq \Gamma = \{w\in \Gamma~|~ \not \exists w'\in \Gamma\mbox{ s.t. }w'\leq w \wedge w\not\leq w'\}$$

Grove shows that for any belief revision operator $\ast$ satisfying the AGM postulates $\mathtt{(R1)}$ - $\mathtt{(R8)}$ and any belief set $B$, there is a system of spheres ${\mathcal{S}_B = \langle W, \leq\rangle}$ such that $w \in Min_\leq W$ iff $w\vDash B$ and $\llbracket B\ast\varphi \rrbracket = Min_\leq \llbracket \varphi\rrbracket$. 

If we take a SOS as representing an agent's belief state, however, we can view a revision operation $\ast$ as an operation that changes a system of spheres $\mathcal{S}_B = \langle W, \leq\rangle$ into a system $\mathcal{S}_{B\ast\varphi} = \langle W, \leq_{\ast\varphi}\rangle$. In that case, we can thus characterise AGM's  revision operators by the \FAITH{} postulate below \cite{nayak2003dynamic}:
\begin{itemize}
\item[] $\mathtt{(Faith)}$ ~~ if $\llbracket \varphi\rrbracket\neq \emptyset$,  $\mathit{Min}_\leq \llbracket \varphi\rrbracket = \mathit{Min}_{\leq_{*\varphi}} W$
\end{itemize}

\textcolor{blue}{Similarly, AGM  introduce a set of postulates $\mathtt{(C1)}$ - $\mathtt{(C8)}$ to define the belief change operation of contraction, which we will omit in this work.  More importantly for us, Grove also shows that for any belief contraction operator $\dot{-}$ satisfying the AGM postulates $\mathtt{(C1)}$ - $\mathtt{(C8)}$ and any belief set $B$, there is a system of spheres ${\mathcal{S}_B = \langle W, \leq\rangle}$ such that $w \in Min_\leq W$ iff $w\vDash B$ and $\llbracket B\dot{-}\varphi \rrbracket = \mathit{Min}_\leq W \cup Min_\leq \llbracket \neg \varphi \rrbracket$. }

\textcolor{blue}{
As before, if we take a SOS as representing an agent's belief state, we can view a contraction operation $\dot{-}$ as an operation that changes a system of spheres $\mathcal{S}_B = \langle W, \leq\rangle$ into a system $\mathcal{S}_{B\dot{-}\varphi} = \langle W, \leq_{\dot{-}\varphi}\rangle$. In that case, we can characterise AGM's  contraction operators by the \GR{} postulate below:
\begin{enumerate}
\item[]\GR{} $Min_{\leq_{\dot{-} \varphi}} W = \mathit{Min}_{\leq} W \cup \mathit{Min}_\leq \llbracket \neg \varphi \rrbracket$.
\end{enumerate}
}

\textcolor{blue}{In the following, we will usually denote by $\leq_{\star \varphi}$ the resulting relation of performing a belief change operation, denoted by some operation symbol $\star$ such as $-$ or $*$, with input $\varphi$ over a relation $\leq$ or, more formally, on a SOS $\mathcal{S} = \langle W,\leq\rangle$.} While AGM \cite{AGM} does not commit to a single logic, in this work, we will focus on propositional classical logic as an object language.

\subsection{Iterated Belief Revision}

AGM belief change says very little about how to change one agent's beliefs repeatedly. In fact, it has been observed that the AGM approach allows some counter-intuitive behaviour in the iterated case.

\textcolor{blue}{One example of such behavior is the following situation \cite{darwiche}: suppose we are introduced to a lady X who sounds smart and looks rich, so we believe that X is smart and X is rich. Moreover, since we profess to no prejudice, we also maintain that X is smart even if found to be poor and, conversely, X is rich even if found to be not smart. Now, we obtain some evidence that X is not smart, and we remain convinced that X is rich. Still, it would be strange for us to say, ``If the evidence turns out false, and X turns out smart after all, we would no longer believe that X is rich''. If we currently believe that X is smart and rich, then evidence first refuting then supporting that X is smart should not change our opinion about X being rich. Strangely, the AGM postulates do permit such a change of opinion.}

\textcolor{blue}{
Belief Change is a continuous process in that new information can always be accommodated into the agent's epistemic state, and successive pieces of information must be incorporated in a principled and coherent manner. 
This fact is implicitly recognized by AGM, as evidenced by postulates $\mathtt{(R7)}$ and $\mathtt{(R8)}$, although the authors do not discuss iteration of change explicitly. }


\textcolor{blue}{To our knowledge, Spohn~\cite{SPO88} was one of the firsts to consider this problem. Spohn~\cite{SPO88} argues for defining belief changes over the agent's belief state, and not only over the agent's held beliefs. Further, the result of a change must be an epistemic state in itself, and this is particularly important, argues the author if we consider repeated epistemic changes. Otherwise, we could not determine the resulting state of a successive change from the initial belief state and the acquired information. To provide an appropriate account for dynamic belief change, the author proposes his semantic framework of Ordinal Conditional Functions.}

\textcolor{blue}{Spohn's work influenced several authors, in particular Darwiche and Pearl~\cite{darwiche}, who propose the further constraints on AGM's belief change operators to account for epistemic consistency across repeated changes.  The authors' proposal can be thought of as encoding Spohn's explicitly defined conditional statements within AGM's framework and establishes constraints (or postulates) that govern how the agent's conditional beliefs are changed. Importantly, the authors present their postulates in terms of both syntactic structures (sets of formulas) and semantic structures (Grove's spheres), which allows us to contrast their postulates with those works based on the AGM framework and with those based on OCF. Here, we focus on the semantic characterisation of these postulates, presented below.}

\begin{enumerate}
\item[]\DP{1} If $w,w'\in \llbracket \varphi\rrbracket$, then $w\leq_{\star\varphi} w'$ iff $w\leq w'$
\item[]$\mathtt{(DP2)}$ If $w,w'\not \in \llbracket \varphi\rrbracket$, then $w\leq_{\star\varphi} w'$ iff $w\leq w'$
\item[]$\mathtt{(DP3)}$ If $w \in \llbracket \varphi\rrbracket$ and $w'\not \in \llbracket \varphi\rrbracket$, then  $w< w'$ implies $w <_{\star\varphi} w'$ 
\item[]$\mathtt{(DP4)}$ If $w \in \llbracket \varphi\rrbracket$ and $w'\not\in \llbracket \varphi\rrbracket$, then $w\leq w'$ implies $w \leq_{\star\varphi} w'$
\end{enumerate}

Nayak et al.~\cite{nayak2003dynamic} argue that the DP postulates are over-permissible, in the sense that they allow revision operators with undesirable properties, \textcolor{blue}{such as  Boutilier's \cite{boutilier1993revision} Natural Revision, which is criticized by Darwiche and Pearl~\cite{darwiche} themselves as having counter-intuitive behaviour.}

Then, Nayak et al.~\cite{nayak2003dynamic} propose the operation of Lexicographic Revision, which can be characterised by the postulates \DP1, \DP{2}, and \REC{} below. The axiom of recalcitrance states that if two pieces of information $\varphi$ and $\psi$ are consistent with each other, then if we obtain the information $\varphi$ and, later, the information $\psi$, there are no grounds to discard $\varphi$. In terms of changes in the relation of a SOS, this can be stated as:

\begin{enumerate}
\item[] \REC{} If $w \in \llbracket \varphi \rrbracket$ and $w'\not \in \llbracket \varphi\rrbracket$, then $w <_{\star\varphi} w'$.
\end{enumerate}

\begin{defn}
Let $\leq ~\subseteq W\times W$ be a \textcolor{blue}{total pre-order} over $W$ and $\varphi$ a propositional formula. The \textit{lexicographic revision} of $\leq$ by information $\varphi$ is the relation $\leq_{*\varphi} \subseteq W\times W$ satisfying postulates \DP{1}, \DP{2} and \REC{}.
\end{defn}

\subsection{Iterated Belief Contraction}

While iterated belief revision has been extensively studied, iterated belief contraction has received far less attention in the literature. AGM~\cite{AGM} show a deep connection between revision and contraction on the single-shot case, by means of the Levi and Harper identities, in the sense that these operations are inter-definable. It is not clear, however, how this connection can be extended to the iterated belief change. In fact, several works on iterated belief contraction stem from trying to establish a connection between iterated revision and contraction. 

Particularly, Chopra et al.~\cite{chopra2008iterated} investigating the role of AGM's recovery principle \cite{AGM} for iterated contraction, propose iterated contraction postulates in the light of Darwich and Pearl's postulates for iterated revision. The authors state their postulates 
in terms of Grove's SOS as:

\begin{enumerate}
\item[]\CR{1} If $w,w' \not\in \llbracket\beta\rrbracket$ then $w\leq w'$ iff $w\leq_{\star\beta} w'$
\item[]\CR{2} If $w,w' \in \llbracket\beta\rrbracket$ then $w\leq w'$ iff $w\leq_{\star\beta} w'$
\item[]\CR{3} If $w \in \llbracket \neg \varphi\rrbracket$ and $w'\in \llbracket \varphi\rrbracket$, then  $w< w'$ implies $w <_{\star\varphi} w'$ 
\item[]\CR{4} If $w \in \llbracket \neg \varphi\rrbracket$ and $w'\in \llbracket \varphi\rrbracket$, then  $w\leq w'$ implies $w \leq_{\star\varphi} w'$
\end{enumerate}

Nayak et al.~\cite{nayak2006taking} propose several iterated contraction operators and analyse their properties.
Among them, we highlight Lexicographic Contraction, which the authors describe as a dual form of Nayak et al.'s Lexicographic Revision.  To define this operator, the authors propose postulate \LC{} which changes the agent's belief state in a manner such that the plausibility attributed to each possible world $w$ is determined solely by their relative position according to the worlds satisfying $\varphi$ (or $\neg \varphi$), if $w$ satisfies $\varphi$ ($\neg \varphi$). This is equivalent to state that an agent maintains a conditional belief $B(\xi ~|~\psi)$\footnote{$B(\xi ~|~\psi)$ stands for the belief that $\psi$ conditionally entails $\xi$, also denoted as $\psi \Rightarrow \xi$ in the literature related to non-monotonic reasoning.} if, and only if, this belief is independent of the agent's attitude towards $\varphi$, i.e. she also holds the belief that $B(\xi~|~\psi \wedge \varphi)$ and $B(\xi~|~\psi \wedge \neg \varphi)$. 

\begin{itemize}
\item[]\LC{} Let $\xi$ be a member of $\{\varphi, \neg\varphi\}$ and $\overline{\xi}$ the other. If $w\vDash \xi$ and $w' \vDash \overline{\xi}$, then $w \leq_{\star \varphi} w'$ iff \textcolor{blue}{there is a chain $w_1, < w_2, \cdots <w_n$ of worlds in $\llbracket\xi\rrbracket$ of maximal length which ends in $w$,  and there is a chain $w'_1, < w'_2, \cdots <w'_k$ of worlds in $\llbracket\overline{\xi}\rrbracket$ which ends in $w'$ and $n\leq k$}.
\end{itemize} 

\textcolor{blue}{
We point out that Lexicographic Contraction, as defined by \cite{nayak2006taking,nayak2007iterated} is a contraction operation constructed on the basis of their proposed Generalised Harper Identity (GHI). By its relation with GHI, lexicographic contraction is a contraction operation based on degrees of plausibility of the possible worlds - encoded in Grove's spheres. This is particularly clear in the fact that the axiomatic characterisation of this operation in \cite{ramachandran2012three} can only be achieved by means of the richer framework of degrees of belief. This connection will be important in defining Lexicographic Contraction for preference models in Section~\ref{sec:dyn}. With the postulate \LC{}, the authors define the operation of lexicographic contraction.}

\begin{defn}
\label{def-lc}
Let $\leq ~\subseteq W\times W$ be a total pre-order over $W$ and $\varphi$ a propositional formula. The \textit{lexicographic contraction} of $\leq$ by information $\varphi$ is the relation $\leq_{-\varphi} \subseteq W\times W$ satisfying postulates \DP{1}, \DP{2} and \LC{}.
\end{defn}

In this work, we explore how the properties (or postulates) discussed in this section can be encoded inside DPL, i.e. how we can guarantee that a given dynamic operator of the logic satisfies one of these postulates. In order to do this, in the following section we introduce DPL based on the work of Girard \cite{girard2008modal} and Souza \cite{souzaphd}.\textcolor{blue}{Lexicographic Revision and Lexicographic Contraction will be explored in this work as examples to illustrate other application of our results.}


%% file: dpl.tex
\section{Dynamic Preference Logic}
\label{sec:dyn}

Preference Logic (or Order Logic, as named by Girard \citeonline{girard2008modal}) is a modal logic complete for the class of transitive and reflexive frames. It has been applied to model a plethora of phenomena in Deontic Logic \cite{liu:deontics}, Logics of Preference~\cite{van2009everything}, Logics of Belief~\cite{BAL08}, etc. 
Dynamic Preference Logic (DPL) \cite{girard2008modal} is the result of ``dynamifying'' Preference Logic, i.e., extending it with dynamic modalities. This logic is one example among several Dynamic Epistemic Logics, and it is particularly interesting for its expressiveness, allowing the study of dynamic phenomena of attitudes such as Beliefs, Obligations, Preferences etc. 

We begin our presentation with the language and semantics of Pre\-fe\-ren\-ce Logic, which we will later ``dynamify''. Let's introduce the language of Preference Logic.
\begin{defn}\label{def:dpl}
Let $P$ be a set of propositional letters. We define the language $\mathcal{L}_{\leq}(P)$ by the following grammar (where $p \in P$): $$
\varphi ::= p ~|~ \neg \varphi ~|~ \varphi \wedge \varphi ~|~ A \varphi ~ | ~ [ \leq ] \varphi~ | ~ [<]\varphi$$
\end{defn}

We will often refer to the language $\mathcal{L}_\leq(P)$ simply as $\mathcal{L}_\leq$, by supposing the set $P$ is fixed. Also, we will denote the language of propositional formulas, i.e., the language removing all modal formulas from $\mathcal{L}_\leq(P)$, by $\mathcal{L}_0(P)$ or simply $\mathcal{L}_0$

\begin{defn}
A \emph{well-founded preference model} is a tuple $M = \langle W, \leq, v\rangle$ where $W$ is a set of possible worlds, $\leq$ is a a reflexive, transitive relation over $W$, s.t its strict part ($<$) is well-founded\footnote{A relation $R\subseteq W\times W$ is said well-founded if there is no infinite descending chains, i.e., for any $\emptyset \neq S\subseteq W$, $Min_\leq S \neq \emptyset$.},  and $v: P \rightarrow 2^W$ a valuation function.
\end{defn}

In such a model, the accessibility relation $\leq$ represents an ordering of the possible worlds according to the preferences of a certain agent. As such, given two possible worlds $w,w' \in W$, we say that $w$ is at least as preferred as $w'$ if $w\leq w'$. 

The interpretation of the formulas over these models is defined as usual.
The $A$ modality is a universal modality\footnote{In this work, we understand the worlds as epistemically possible worlds, not metaphysically possible. While formally this difference is irrelevant, philosophically it is of importance.} satisfied iff all worlds in the model satisfy its argument. The $[\leq]$ modality is a box modality on the accessibility order $\leq$. The $[<]$ modality is the strict variant of $[\leq]$. They are interpreted as: \textcolor{blue}{$$\begin{array}{lll}
M,w \vDash p &\mbox{ iff } &w\in v(p)\\
M,w \vDash \neg \varphi &\mbox{ iff } &M,w \not\vDash \varphi\\
M,w \vDash \varphi \wedge \psi  &\mbox{ iff } &M,w \vDash \varphi \mbox{ and } M,w \vDash \psi\\
M,w \vDash A\varphi &\mbox{ iff } &\forall w' \in W:\, M,w'\vDash\varphi\\
M,w \vDash [\leq]\varphi & \mbox{ iff }&\forall w' \in W:\mbox{ if } w'\leq w \mbox{ then } M,w' \vDash  \varphi\\
M,w \vDash [<]\varphi & \mbox{ iff }&\forall w' \in W:\mbox{ if } w'< w \mbox{ then } M,w' \vDash \varphi
\end{array}$$}

As usual, we will refer as $E\varphi$ to the formula $\neg A \neg \varphi$, meaning `\textit{it is possibly true that} $\varphi$', and as $\langle \leq \rangle \varphi$ ($\langle < \rangle \varphi$) to the formula $\neg [\leq]\neg\varphi$ ($\neg [<]\neg\varphi$), meaning `\textit{in a possible situation at least as (strictly more) preferable as the current one,} $\varphi$ \textit{holds},' as commonly done in modal logic.

For  simplicity, in this work, we will refer to well-founded preference models only as preference models. Notice, however, that in the literature, e.g., \cite{girard2008modal}, preference models need not be well-founded.

Given a preference model $M$ and a formula $\varphi$, we use the notation $\llbracket \varphi\rrbracket_M$, as also defined for systems of spheres,  to denote the set of all the worlds in $M$ satisfying $\varphi$, or only $\llbracket \varphi\rrbracket$ when the model is clear from the context. The notation $Min_\leq X$ denotes the `most preferred worlds in $X$' in the model, i.e., the minimal elements in a set of possible worlds $X$, according to the (pre-)order $\leq$. 

As the concept of most preferred worlds satisfying a given formula $\varphi$ will be of great use in modelling different notions of belief (and different belief change operations) in this logic, we define a formula encompassing this exact concept: $$\mu\varphi \equiv_{def} \varphi \wedge \neg \langle < \rangle \varphi.$$ 

It is easy to see that $\mu \varphi$ defines exactly the set of minimal worlds satisfying $\varphi$.

\begin{prop}\label{def:mu}\cite{liu2011reasoning}
Let $M=\langle W, \leq, v\rangle$ be a preference model and $\varphi\in \mathcal{L}_\leq$ a formula. For any $w\in W$\!, $M,w\vDash \mu \varphi$ iff $w \in Min_\leq \llbracket \varphi\rrbracket_M$.
\end{prop}

Notice that Preference Logic is highly expressive. It is well known \cite{boutilier1994conditional,girard2008modal,liu2011reasoning,souzaphd} that we can define unconditional and conditional preferences, i.e. the notions of `\textit{the agent prefers that} $\varphi$' and `\textit{in the case of} $\psi$, \textit{the agent would prefer that} $\varphi$'. In the context of Belief Change, we can interpret this preference as belief and, as such, we can encode in this logic, as previously done by \cite{boutilier1994conditional,liu2011reasoning}, the notion of conditional belief as $$B(\varphi | \psi) \equiv_{\mathit{def}} A ( \mu \psi \rightarrow \varphi)$$ and unconditional belief as $$B(\varphi)\equiv_{\mathit{def}	} B(\varphi | \top).$$

\textcolor{blue}{These encodings preserve the usual notion of conditional belief as the ``\textit{the most plausible worlds satisfying }$\varphi$ \textit{also satisfy }$\psi$''. 
\begin{corolary}
Let $M=\langle W, \leq, v\rangle$ be a preference model, $w\in W$ be a possible world, and $\varphi,\psi\in \mathcal{L}_\leq$ be preference formulas.
$$\begin{array}{lll}
M,w\vDash B(\psi~|~\varphi) & \mbox{iff} & Min_\leq \llbracket \varphi \rrbracket \subseteq \llbracket \psi\rrbracket
\end{array}$$ 
\end{corolary}}

Souza \cite{souzaphd} provided the axiomatisation depicted in Figure~\ref{fig:PLaxiom} below for the logic and showed that it is  weak-complete \cite{blackburn2006handbook} for Preference Logic restricted to well-founded models.

\begin{figure}[ht]
\textcolor{blue}{\caption{Axiomatization for Preference Logic}
\label{fig:PLaxiom}
$$
\begin{array}{ll}
\mathbf{CP} & \mbox{All axioms from Classical Propositional Logic}\\
\mathbf{K}_\leq:& [\leq](\varphi\rightarrow \psi)\rightarrow ([\leq]\varphi \rightarrow [\leq]\psi)\\
\mathbf{T}_\leq: &[\leq]\varphi\rightarrow \varphi\\
\mathbf{4}_\leq: & [\leq]\varphi \rightarrow [\leq][\leq]\varphi\\
\mathbf{K}_<:& [<](\varphi\rightarrow \psi)\rightarrow ([<]\varphi \rightarrow [<]\psi)\\
\mathbf{W}_<:& [<]([<]\varphi\rightarrow \varphi)\rightarrow [<]\varphi\\
<\leq_1:&[\leq]\varphi \rightarrow [<]\varphi\\
<\leq_2:&[<]\varphi \rightarrow [<][\leq]\varphi\\
<\leq_3:&[<]\varphi \rightarrow [\leq][<]\varphi\\
<\leq_4:& [\leq]([\leq]\varphi \vee \psi) \wedge [<]\psi \rightarrow \varphi \vee [\leq]\psi\\
\mathbf{K}_A: & A(\varphi\rightarrow \psi)\rightarrow (A\varphi \rightarrow A \psi)\\
\mathbf{T}_A: &A\varphi\rightarrow \varphi\\
\mathbf{4}_A: & A\varphi \rightarrow AA\varphi\\
\mathbf{B}_A:&\varphi \rightarrow A \neg A \neg\varphi\\
A\leq:& A\varphi \rightarrow [\leq]\varphi\\
\\
(\mathit{Necessitation})& \vdash\varphi ~\Rightarrow ~ \vdash\Box\varphi,~ \mbox{ with }\Box\in\{[\leq],[<], A\}\\
(\mathit{Modus~Ponens}) &  \Gamma \vdash\varphi \mbox{ and }\Gamma \vdash\varphi\rightarrow \psi ~~\Rightarrow~~  \Gamma \vdash\psi
\end{array}
$$
}
\end{figure}

\subsection{A Logic of Iterated Belief Change}
\label{sec:dynamic}

We will now reintroduce some of the belief change operations presented in Section~\ref{sec:ibr} as transformations of preference models.With this, we will dynamify Preference Logic by introducing dynamic modalities in the language to represent the execution of a belief change operation in the agent's epistemic state. 

We start with lexicographic revision of an epistemic state by a formula $\varphi$, which consists of making each world satisfying $\varphi$ strictly more preferable than those not satisfying it, while maintaining the order otherwise. 
\begin{defn}
\label{def:RU} \cite{girard2008modal}
Let $M = \langle W, \leq, v\rangle$ be a preference model and $\varphi \in \mathcal{L}_0$. We say the model  $M_{\Uparrow\varphi} = \langle W, \leq_{\Uparrow\varphi},  v\rangle$ is the result of the lexicographic revision of $M$ by $\varphi$,  where 
$$ w \leq_{\Uparrow \varphi} w' \mbox{ iff } \begin{cases} 
w \leq w' \mbox{ and }w,w' \in \llbracket \varphi \rrbracket\mbox{, or}\\
w \leq w' \mbox{ and if }w,w' \not\in \llbracket \varphi \rrbracket\mbox{, or}\\
w\in \llbracket \varphi \rrbracket  \mbox{ and }w' \not\in \llbracket \varphi \rrbracket
\end{cases}$$
\end{defn}

We can now introduce the modality $[\Uparrow\!\!\varphi]$ in the language of $\mathcal{L}_\leq$, where $[\Uparrow\!\varphi]\psi$ is read as ``after the lexicographic revision by $\varphi$, $\psi$ holds''.

\begin{defn}\label{def:logiclexrev}
We define the language of Preference Logic extended with Lexicographic Revision, denoted by $\mathcal{L}_\leq(\Uparrow)$, as the language constituted of all formulas of  $\mathcal{L}_\leq$ as well as any formula $[\Uparrow \varphi]\psi$ s.t. $\varphi \in \mathcal{L}_0$ and $\psi \in \mathcal{L}_\leq(\Uparrow)$. More yet, let $M = \langle W, \leq, v\rangle$ be a preference model, $w\in W$ and $\varphi$ a formula of $\mathcal{L}_0$
$$M,w\vDash [\Uparrow\!\!\varphi]\psi \qquad \mbox{if} \qquad M_{\Uparrow\varphi},w\vDash \psi$$
\end{defn}
.

Liu~\cite{liu2011reasoning} has shown that Preference Logic extended with Lexicographic Revision is completely axiomatised by the axioms for Preference Logic extended by the reduction axioms and rules depicted in Figure~\ref{fig:axiomRU}. \textcolor{blue}{The authors obtain such axiomatization using a Propositional Dynamic Logic (PDL) \cite{fischer1979propositional} codification of Lexicographic Revision to derive the axioms \cite{pref}.}

\textcolor{blue}{A PDL program is a regular expression over some set of symbols representing basic relations in some interpretation set.  Fisher and Laudner~\cite{fischer1979propositional} introduced PDL programs in their propositional variation of Pratt's \cite{pratt1976semantical} Dynamic Logic to represent computer programs. Van Benthem and Liu~\cite{pref} have used them to encode belief change operators.  For space constraints, we do not include in this paper a detailed explanation about PDL programs. We refer the reader to \cite{harel2001dynamic} for more details.  }

\begin{figure}
\centering
\caption{Reduction axioms for the Lexicographic Revision}
\label{fig:axiomRU}
\vspace{0.1cm}
$\begin{array}{lll}
{}[\Uparrow\varphi]p &\leftrightarrow & p\\
{}[\Uparrow\varphi]\neg \psi&\leftrightarrow & \neg[\Uparrow\varphi]\psi\\
{}[\Uparrow\varphi](\psi\wedge\xi)&\leftrightarrow & [\Uparrow\varphi]\psi \wedge [\Uparrow\varphi] \xi\\
{}[\Uparrow\varphi]A\psi &\leftrightarrow & A( [\Uparrow\varphi] \psi)\\
{}[\Uparrow\varphi][\leq]\psi &\leftrightarrow & \varphi \rightarrow [\leq](\varphi \rightarrow [\Uparrow\varphi]\psi)\wedge\\
   && \neg\varphi\rightarrow (A(\varphi \rightarrow [\Uparrow\varphi]\psi) \wedge [\leq](\neg\varphi \rightarrow [\Uparrow\varphi]\psi)) \\
{}[\Uparrow\varphi][<]\psi &\leftrightarrow & \varphi \rightarrow [<](\varphi \rightarrow [\Uparrow\varphi]\psi)\wedge \\
   &&\neg\varphi\rightarrow (A( \varphi \rightarrow [\Uparrow\varphi]\psi) \wedge [<](\neg\varphi \rightarrow [\Uparrow\varphi]\psi)) \\
\vdash\psi & \Rightarrow & \vdash[\Uparrow \varphi]\psi \\
\end{array}$
\end{figure}

As for Lexicographic Revision, we can provide axiomatisations for Preference Logic extended with other belief change operations using their codification in PDL. However, since some operations cannot  be represented as PDL programs, this strategy is not viable for all known belief change operations. To give an example of such an operation, let's define Lexicographic Contraction as a transformation on preference models. Later on, in Section~\ref{sec:iter}, we will provide a method to derive an axiomatization for Preference Logic extended with Lexicographic Contraction that does not relies on PDL encoding of belief change operators.

To encode Lexicographic Contraction in DPL, we need to be able to encode the notion of a chain of worlds of a given size $i$ all of which satisfy a formula $\varphi$, \textcolor{blue}{a notion connected to that of degree of belief}, denoted by $dg_\varphi(i)$: 

\begin{defn}
Let $M = \langle W, \leq, v\rangle$ be a preference model, $\varphi$ a formula of $\mathcal{L}_0$, and $i\in \mathbb{N}$ a natural number. We define the formula $dg_\varphi(i)$ as:
$$
dg_\varphi(i) = \begin{cases}
                  \varphi & \mbox{if } i=1\\
                   \varphi \wedge \langle < \rangle \, dg_\varphi(i\!-\!1)& \mbox{if } i> 1\\
				\end{cases}
$$
\end{defn}

The maximal natural number $i$, such that $dg_\varphi(i)$ is satisfied by some world $w$ in a model $M$,  called the implausibility degree of $w$ in $M$,  is the size of the maximal chain of $\varphi$-worlds ending with the world $w$.

\begin{lema}\label{lem:dg}\cite{souzaphd}
Let $M = \langle W, \leq, v\rangle$ be a preference model, $\varphi$ a formula of $\mathcal{L}_0$ and $w\in W$. $M,w \vDash dg_\varphi(i), i>1$ iff there is a chain of worlds of $w_1 < w_2 \ldots < w_i$, such that, $w_j \in \llbracket \varphi\rrbracket$, for all $j=1..i$, $w_1 \in Min_\leq \llbracket \varphi\rrbracket$, and $w_i = w$. Also, if $M,w \vDash \mu \, dg_\varphi(i)$, \textcolor{blue}{then there is no other chain of $\varphi$-worlds ending in $w$ of greater size.}
\end{lema}

Notice that the operation of lexicographic contraction, as characterised by Ramachandran et al.~\cite{ramachandran2012three}, is not closed for preference models (see Definition~\ref{def-lc}). The reason for this is that their characterisation assumes the pre-order over the possible worlds is total. In fact, in the case of preference models, the axioms \DP{1} and \LC{} are incompatible, since they may result in loss of transitivity. As such, we propose the following modification for preference models.

\begin{itemize}
\item[]\LC{'}: Let $\xi,\chi $ be members of $\{\varphi, \neg\varphi\}$ - not necessarily distinct. If $w\vDash \xi$ and $w' \vDash \chi$, then $w \leq_{\star \varphi} w'$ iff the maximal length of a chain of $\xi$-worlds which ends in $w$ is smaller or equal to the maximal length of a chain of $\chi$-worlds which ends in $w'$.
\end{itemize} 

\textcolor{blue}{Notice that postulate \LC{'} is rather strong compared to the original \LC{} postulate.  The reason for this is that \LC{'} is  an amalgamation of three postulates,  namely (LC),  and the modifications of (DP1), (DP2) below.}

\begin{itemize}
\item[] \textcolor{blue}{\DP{1'} If $w,w'\in \llbracket \varphi\rrbracket$, then $w \leq_{\star \varphi} w'$ iff the  maximal length of a chain of $\varphi$-worlds which ends in $w$ is smaller or equal to the maximal length of a chain of $\varphi$-worlds which ends in $w'$.}
\item[] \textcolor{blue}{\DP{2'} If $w,w'\not\in \llbracket \varphi\rrbracket$, then $w \leq_{\star \varphi} w'$ iff the  maximal length of a chain of $\neg\varphi$-worlds which ends in $w$ is smaller or equal to the  maximal length of a chain of $\neg\varphi$-worlds which ends in $w'$.}
\end{itemize}

\textcolor{blue}{
From a philosophical point of view, lexicographic contraction is a contraction operation based on degrees of plausibility of the possible worlds - encoded in Grove's spheres. As such, based on degrees of plausibility alone, \DP{1'} and \DP{2'} (and thus \LC{'}) are justified modifications for preference models, since they imply that in the resulting epistemic state comparability between the worlds is determined by their degree of plausibility.}

With postulate \LC{'} we can define a lexicographic contraction operator over preference models  - which coincides with Ramachandran et al.'s~\cite{ramachandran2012three} on Grove models.

\begin{defn}\label{def:lexi}
Let $M = \langle W, \leq, v\rangle$ be a preference model and $\varphi$ a formula of $\mathcal{L}_0$. We say the model  $M_{ \Downarrow \varphi} = \langle W, \leq_{\Downarrow \varphi},  v\rangle$ is the lexicographic contraction of $M$ by $\varphi$,  where: 

{
$w\leq_{\Downarrow \varphi} w' \mbox{ iff } 
\begin{cases}
     w\in \llbracket \mu dg_\varphi(i)\rrbracket\mbox{ and } w'\in \llbracket \mu dg_{ \varphi}(j)\rrbracket & i\leq j \mbox{, or }\\
     w\in \llbracket \mu dg_{\neg\varphi}(i)\rrbracket\mbox{ and } w'\in \llbracket \mu dg_{\neg \varphi}(j)\rrbracket & i\leq j  \mbox{, or }\\
     w\in \llbracket \mu dg_\varphi(i)\rrbracket\mbox{ and } w'\in \llbracket \mu dg_{\neg \varphi}(j)\rrbracket & i\leq j  \mbox{, or }\\
w\in \llbracket \mu dg_{\neg \varphi}(i)\rrbracket\mbox{ and } w'\in \llbracket \mu dg_{\varphi}(j)\rrbracket & i\leq j \end{cases}$
}
\end{defn}

Again, we can introduce the modality $[\Downarrow\!\!\varphi]$ in the language of $\mathcal{L}_\leq$, where $[\Downarrow\!\varphi]\psi$ is read as ``after the lexicographic contraction by $\varphi$, $\psi$ holds''. 

\begin{defn}\label{def:logiclexcont}
We define the language of Preference Logic extended with Lexicographic Contraction, denoted by $\mathcal{L}_\leq(\Downarrow)$, as the language constituted of all formulas of  $\mathcal{L}_\leq$ as well as any formula $[\Downarrow \varphi]\psi$ s.t. $\varphi \in \mathcal{L}_0$ and $\psi \in \mathcal{L}_\leq(\Downarrow)$. More yet, let $M = \langle W, \leq, v\rangle$ be a preference model, $w\in W$ and $\varphi$ a formula of $\mathcal{L}_0$
$$M,w\vDash [\Downarrow\!\!\varphi]\psi \qquad \mbox{if} \qquad M_{\Downarrow\varphi},w\vDash \psi$$
\end{defn}

While the axiomatisations provided in the literature for the other belief change operators were constructed by means of PDL representations of these operations, this technique cannot be applied to Lexicographic Contraction due to the simple fact that this operation is not regular. 

\begin{fact}\label{fact:13pdl}
Lexicographic Contraction cannot be encoded by means of PDL programs
\end{fact}

To provide an axiomatisation for this operation, Souza~\cite{souzaphd} shows that if we restrict the logic's semantics to consider only models with \textcolor{blue}{chains of a maximum finite size then an axiomatisation} can be achieved by means of the PDL representation of Lexicographic Contraction. Later in Section~\ref{sec:iter}, we will employ a new technique, not based in PDL, to derive an axiomatisation for the extended language, which does not require such a drastic restriction in its semantics.

\subsection{A General Notion of Dynamic Operator}
We define a dynamic operation on a preference model as any operation that takes a preference model and a propositional formula and changes only the preference relation of the model. Let $\mathit{Mod}(\mathcal{L}_\leq)$ denote the class of all preference models for the language $\mathcal{L}_\leq$, with the syntax given in Definition~\ref{def:dpl}.

\begin{defn}\label{def:op}
Let $\star: \mathit{Mod}(\mathcal{L}_\leq)\times \mathcal{L}_0 \rightarrow \mathit{Mod}(\mathcal{L}_\leq)$, we say $\star$ is a dynamic operator on preference models if for any preference model $M= \langle W,\leq,v\rangle$ and propositional formula $\varphi\in \mathcal{L}_0$, $\star(M,\varphi) = \langle W_\star, \leq_\star, v_\star\rangle$\textcolor{blue}{, with $W_\star =W$ and $v_\star = v$}.
\end{defn}

We limited our dynamic operators not to change the set of possible worlds \textcolor{blue}{or valuations}. This limitation is justified by the fact that we are considering belief changing operators, i.e., mental actions which change the plausibility the agent attributes to each epistemically possible world, not creating any new epistemic certainty (knowledge) \textcolor{blue}{nor having effects on the world}. 

Given a dynamic operator $\star$, we extend the language $\mathcal{L}_\leq$ with formulas $[\star \varphi]\xi$. 

\begin{defn}
Let $\star$ be a symbol for dynamic operators. We define the language $\mathcal{L}_\leq (\star)$ as the smallest set containing $\mathcal{L}_\leq$ and all formulas $[\star \varphi]\xi$, with $\varphi \in \mathcal{L}_0$ and $\xi \in \mathcal{L}_\leq (\star)$.
\end{defn}

As for the case of DPL of Lexicographic Revision and DPL of Lexicographic Contraction (c.f. Definitions~\ref{def:logiclexrev} and \ref{def:logiclexcont}) presented before, the formulas of $\mathcal{L}_\leq (\star)$ must be interpreted over a preference model and a dynamic operator that describes the changes in the model. As such, we will introduce the notion of dynamic preference model containing these two ingredients.

\begin{defn}
A \textit{dynamic preference model}, or simply a \textit{dynamic model}, is a tuple $\langle M, \star\rangle$, where $M \in \mathit{Mod}(\mathcal{L}_\leq)$ is a preference model and $\star: \mathit{Mod}(\mathcal{L}_\leq)\times \mathcal{L}_0 \rightarrow \mathit{Mod}(\mathcal{L}_\leq)$ is a dynamic operator.
\end{defn}

In the definition above,  $\star$ is used both as a symbol  denoting a dynamic operator, seem as a function,  and also as a symbol in the object logic language. With that, we can define how formulas in the language $\mathcal{L}_\leq(\star)$ are interpreted.

\begin{defn}\label{def:semantics}
Let $D = \langle M, \star\rangle$ be a dynamic model, with $M = \langle W, \leq, v\rangle$,  $w\in W$ be a possible world, and $\xi \in \mathcal{L}_\leq(\star)$ be a dynamic preference formula. We define the satisfiability of $\xi$ by $w$ in $D$, denoted by $D,w \vDash \xi$, as follows:
$$\begin{array}{ll}
D, w \vDash p & \mbox{if }w \in v(p)\\
D, w \vDash\varphi \wedge \psi &\mbox{if } D,w\vDash\varphi \mbox{ and } D, w \vDash \psi\\
D, w \vDash \neg\varphi & \mbox{if }D,w\not \vDash \varphi\\
D, w \vDash A \varphi & \mbox{if, for all }w'\in W: D, w'  \vDash \varphi\\
D, w \vDash [\leq] \varphi & \mbox{if, for all }w'\in W\mbox{ s.t. }w'\leq w: D, w'  \vDash \varphi\\
D, w \vDash [<] \varphi & \mbox{if, for all }w'\in W\mbox{ s.t. }w'< w: D, w'  \vDash \varphi\\
D,w\vDash [\star \varphi]\psi & \mbox{if } D',w \vDash \psi \mbox{, where }D'=\langle \star(M,\varphi), \star\rangle 
\end{array}$$
As usual, we say $\xi$ is valid in $D$, denoted by $D\vDash \xi$, if for all $w\in W$, $D,w\vDash \xi$, and that $\xi$ is valid, denoted $\vDash \xi$, if it is valid for any dynamic preference model $D$.
\end{defn}

Since in this work we investigate axiomatisations that characterise classes of dynamic operators, we will need to define the notion of a formula been valid on a class of dynamic models. For that, we define the notion of a class of dynamic models, which will be used in Section~\ref{sec:iter}, similarly to how classes of frames are connected to modal axioms in correspondence theory for Modal Logic \cite{blackburn2006handbook}.  

\begin{defn}
Let $\mathfrak{M} \subseteq \textit{Mod}(\mathcal{L}_\leq)$ be a class of preference models and $\mathfrak{C}$  a class of dynamic operators closed over $\mathfrak{M}$, i.e., for any $\varphi\in \mathcal{L}_0$, $\star(\mathfrak{M},\varphi)\subseteq \mathfrak{M}$ for each $\star \in \mathfrak{C}$. We denote by $\mathfrak{D}=\langle \mathfrak{M},\mathfrak{C}\rangle$ the class of dynamic models $\langle M,\star \rangle$ s.t. $M\in \mathfrak{M}$ and $\star \in \mathfrak{C}$. 
\end{defn}

If $\mathfrak{D}$ is a class of dynamic models and $\xi\in \mathcal{L}_\leq(\star)$ is a dynamic preference formula, we will often say $\xi$ is satisfiable in $\mathfrak{D}$ if there is some dynamic model $D\in \mathfrak{D}$ in which $\xi$ is satisfiable. Similarly, we will say that $\xi$ is valid in $\mathfrak{D}$ if for any $D \in \mathfrak{D}$, $D\vDash \xi$.

%% file: ibc.tex
\section{Iterated Belief Change and DPL}
\label{sec:iter}

In this section, we investigate the relationship between the postulates satisfied by
iterated belief change operators discussed in Section~\ref{sec:ibr} and the axioms satisfied in DPL using these operators. In other words, given an iterated belief change postulate $P$, we investigate which validities are induced in the logic of $\mathcal{L}_\leq(\star)$ if we consider only classes of models in which the dynamic operators satisfy postulate $P$. More yet, we study how to characterise this postulate in DPL, i.e., which axioms should be introduced in the axiomatization of DPL that imply that the models of this logic must satisfy $P$. 

We examine Darwiche and Pearl's \DP{1}-\DP{4}  \cite{darwiche}, as well as Nayak et al's \REC{} \cite{nayak2003dynamic}, for iterated belief revision,  and Chopra et al's \CR{1}-\CR{4} \cite{chopra2008iterated} and Ramachandran et al's \LC{} \cite{ramachandran2012three} for iterated belief contraction. Other postulates in the literature can be easily encoded in the same way based on our results. The proofs of the results presented in this section are given in the Appendix.

First, it is easy to see that for any dynamic operator $\star$, the extended logic must satisfy some basic principles.

\begin{prop}\label{prop:BasicAxiom}
The following reduction axioms and rule are valid, for any propositional symbol $p\in P$, propositional formula $\varphi \in \mathcal{L}_0$ and formula $\xi \in \mathcal{L}_\leq(\star)$.
\vspace{-0.2cm}
$$\begin{array}{lcl}
{}[\star\varphi]p &\leftrightarrow& p\\
{}[\star\varphi](\xi_1 \wedge \xi_2)&\leftrightarrow& [\star\varphi]\xi_1 \wedge [\star\varphi]\xi_2\\
{}[\star\varphi]\neg \xi&\leftrightarrow&\neg [\star\varphi]\xi\\
{}[\star\varphi]A \xi &\leftrightarrow& A [\star\varphi] \xi\\
{}\vdash \xi & \Rightarrow & \vdash [\star \varphi]\xi
\end{array}$$
\end{prop}

As in Souza's axiomatisation for DPL \cite{souzaphd}, we do not require Uniform Substitution as a rule in our proof systems. This is an important characteristic of our proof systems, and quite common in dynamic epistemic logics \cite{herzig2017dynamic}. If Uniform Substitution was included in the system, it would be possible to derive $\vdash [\star \varphi]\xi \leftrightarrow \xi$ for any $\xi \in \mathcal{L}_\leq(\star)$ and $\varphi \in \mathcal{L}_0$, i.e., the dynamic modality would be frivolous. Without Uniform Substitution, $[\star \varphi]\xi \leftrightarrow \xi$ can only be derived by the axioms in Proposition~\ref{prop:BasicAxiom} if $\xi \in \mathcal{L}_0$, i.e., $\xi$ is a propositional formula. This is consistent with the interpretation that our dynamic operators are mental actions and, thus, do not change the propositional (or ontic) properties of the worlds in a model.

We wish to investigate which properties are induced in the dynamic logic, given the postulates satisfied by a given dynamic operator $\star$. \textcolor{blue}{First, we study the characterisation of iterated belief revision postulates in DPL, then we extend this study to iterated belief contraction postulates. Further, we show how these characterisations can used to derive axiomatisations for a logic defined over classes of models satisfying a set of iterated belief change postulates.}

%
\subsection{Postulates for Iterated Belief Revision in DPL}
\label{subsec:IBRinDPL}

First, let us consider the basic postulate defining belief revision operations, namely the postulate of $\mathtt{(Faith)}$ which indicates that the dynamic operation $\star$ is an AGM belief revision operator.

\begin{prop}
\label{prop:Faith} 
Let $\mathfrak{C}$ be a class of dynamic operators $\star$ satisfying $\mathtt{(Faith)}$ and $\mathfrak{M}\subseteq \mathit{Mod}(\mathcal{L}_\leq)$ be a class of preference models. The following axiom schema is valid in $\langle \mathfrak{M}, \mathfrak{C}\rangle$, for any propositional formula $\varphi \in \mathcal{L}_0$.
\vspace{-0.2cm}
$$
\begin{array}{lcl}
E\varphi  & \rightarrow  & \mu\varphi \leftrightarrow [\star\varphi]\mu\top\\
\end{array}
$$
\end{prop}

Postulate \FAITH{} states that after a revision by some information $\varphi$, the minimal elements of the agent's belief state are exactly the most plausible $\varphi$-states. As such, the axiom schema in Proposition~\ref{prop:Faith} states that, if a proposition $\varphi$ is satisfiable in a model $M$, for any minimal $\varphi$-world $w$ in $M$, it holds that after change by the dynamic operator $\star$, $w$ is a minimal world of the resulting model, and vice-versa. 

We can also show that the axiom schema of  Proposition \ref{prop:Faith} completely characterises the postulate \FAITH{} in DPL. In the results below, we will often denote a class of dynamic model $\langle \mathfrak{M},\{\star\}\rangle$ simply by $\langle \mathfrak{M},\star\rangle$.

\begin{prop}
\label{prop:charFaith} 
Let $\mathfrak{D}=\langle \mathfrak{M}, \star\rangle$ be a class of dynamic models. The axiom schema in Proposition~\ref{prop:Faith} is valid in $\mathfrak{D}$, for any propositional formula $\varphi \in \mathcal{L}_0$ iff 
for each $M = \langle W, \leq, v\rangle \in \mathfrak{M}$, 
s.t. $\star(M, \varphi) = \langle W, \leq_{\star \varphi}, v\rangle$, if $\llbracket \varphi\rrbracket\neq \emptyset$, 
then  $Min_\leq \llbracket \varphi \rrbracket = Min_{\leq_{\star \varphi}} W$,
i.e., $\star$ satisfies \FAITH{} in $\mathfrak{M}$.
\end{prop}

While \FAITH{} is the basic postulate for AGM revision, it does not imply any constraints on the iterated properties of the operation. To characterise iterated belief revision in DPL, we will consider the logic characterisation of Darwiche and Pearl's \DP{1}-\DP{4}  \cite{darwiche} postulates, as well as Nayak et al.'s \REC{} \cite{nayak2003dynamic}.
%

\paragraph{\textcolor{blue}{Postulate DP1}}
\textcolor{blue}{Let's start with  postulate \DP{1}, as presented in Section~\ref{sec:ibr}.
\begin{itemize}
\item[] \DP{1} If $w,w'\in \llbracket \varphi\rrbracket$, then $w\leq_{\star\varphi} w'$ iff $w\leq w'$
\end{itemize}
It is easy to see that this property implies the following validities in the logic defined by the operators satisfying it.}
 
\begin{prop}
\label{prop:CR1}
Let $\mathfrak{C}$ be a class of dynamic operators $\star$ satisfying \DP{1} and $\mathfrak{M}\subseteq \mathit{Mod}(\mathcal{L}_\leq)$ be a class of preference models. The following axiom schemata is valid in $\langle \mathfrak{M}, \mathfrak{C}\rangle$ for any $\varphi \in \mathcal{L}_0$ and $\xi \in \mathcal{L}_\leq(\star)$.
\vspace{-0.2cm}
$$
\begin{array}{lcl}
{}[\star\varphi][\leq]\xi &\rightarrow& (\varphi \rightarrow [\leq](\varphi \rightarrow [\star\varphi] \xi))\\
{}[\star\varphi][<]\xi &\rightarrow& (\varphi \rightarrow [<](\varphi \rightarrow [\star\varphi] \xi))\\
{}[\leq][\star\varphi]\xi &\rightarrow& (\varphi \rightarrow [\star\varphi][\leq](\varphi \rightarrow \xi))\\
{}[<][\star\varphi]\xi &\rightarrow& (\varphi \rightarrow [\star\varphi][<](\varphi \rightarrow \xi))
\end{array}$$
\end{prop}

Notice that \DP1 establishes a bidirectional relationship between the preference relations $\leq$ and $\leq_{\star\varphi}$ by a double implication. Each side of this implication is captured in the axiom schemata in Proposition~\ref{prop:CR1}, where the first axiom states that if there is two $\varphi$-world $w,w'$ s.t. $w'\leq_{\star\varphi} w$,  and $w'$ satisfies $\xi$, then there is some $\varphi$-world $w''$ satisfying $[\star \varphi]\xi$ and $w''\leq w$. This first axiom  of Proposition~\ref{prop:CR1} generalises the requirement that if $w \leq_{\star\varphi} w'$ then $w \leq w'$ in \DP{1},  
while the third axiom  states the opposite direction of this relation. The second and fourth axioms are variations that express the changes in the strict part $<$ of the preference relation.

Proposition~\ref{prop:CR1} provides a representation of the postulate \DP{1} as a DPL axiom schemata induced,  in the extended logic,  by operations $\star$ that satisfy this postulate. 
However, The logical characterisation of the postulate, i.e., that if the logic satisfies a certain set of axioms, then the dynamic operator must satisfy \DP{1}, as established for \FAITH{} in Proposition~\ref{prop:charFaith}, cannot be achieved in DPL. The reason for this is that the language is not expressive enough to distinguish every world in the model. As such, there may be worlds in a model that are ``modally equivalent'' (m.e.) \cite{blackburn2006handbook}, in the sense that they satisfy exactly the same formulas, and there is no way to express in the logic any relation that differentiates these worlds. As such, it is easy to construct dynamic operators that are equivalent, in the sense that the logics generated by them are the same.

\begin{fact}\label{fact:cr1}
There are two dynamic operators $\star,\ast:\mathit{Mod}(\mathcal{L}_\leq)\times \mathcal{L}_0 \rightarrow \mathit{Mod}(\mathcal{L}_\leq)$ s.t. for any class of preference models $\mathfrak{M}$, s.t. both $\star$ and $\ast$ are closed over $\mathfrak{M}$, and formula $\xi \in \mathcal{L}_\leq(\star)$, $\xi$ is satisfiable in $\langle \mathfrak{M}, \star\rangle$ iff $\xi$ is satisfiable in $\langle \mathfrak{M}, \ast\rangle$, but $\star$ does not satisfies \DP{1} while $\ast$ does.
\end{fact}

It is easy to see from the proof of Fact~\ref{fact:cr1} (c.f. the Appendix) and from the axioms presented in Proposition~\ref{prop:CR1} that some dynamic operators may fail to satisfy \DP{1} and yet preserve all conditional beliefs regarding the new information $\varphi$. 
\textcolor{blue}{In fact, they satisfy Darwiche and Pearl's original formulation of postulate \DP{1} based on conditional beliefs \cite{darwiche}.}

\textcolor{blue}{\begin{prop}\label{prop:dp1syn}
Let $\mathfrak{D} = \langle \mathfrak{M}, \star\rangle$ be a class of dynamic models satisfying the axiom schemata in Proposition~\ref{prop:CR1}. For any propositional formulas $\varphi, \psi \in \mathcal{L}_0$ s.t. ${\mathfrak{D}\vDash \varphi \rightarrow \psi}$ and any dynamic preference formula $\xi \in \mathcal{L}_\leq(\star)$, it holds that
$$\mathfrak{D}\vDash [\star \psi]B(\xi~|~\varphi) \leftrightarrow B([\star \psi] \xi ~|~\varphi)$$
\end{prop}} 

The reason for this is that the semantic formulation of \DP{1} is rooted in the identities of the worlds in the model, not in the information they hold. More accurately, \DP{1} assumes that for every pair of worlds $w,w'$ in the model, there is some proposition $\xi$ that only one of them satisfies. As such, Fact~\ref{fact:cr1} points out that the DP postulates based on Grove's model need to be generalised to our models. We then define the notion of DP1-compliance in a certain class of models, meaning that a  dynamic operator $\star$ behaves as to preserve the agent's beliefs conditioned to the new information $\varphi$.

\begin{defn}\label{def:dp1plus}
Let $\mathfrak{M}\subseteq \mathit{Mod}(\mathcal{L}_\leq)$ be a class of preference models, and let $\star: \mathit{Mod}(\mathcal{L}_\leq)\times \mathcal{L}_0 \rightarrow \mathit{Mod}(\mathcal{L}_\leq)$ be a dynamic operator. We say $\star$ is DP1-compliant in regards to $\mathfrak{M}$, or $\mathfrak{M}$-DP1-compliant, if, for any preference model ${M = \langle W, \leq, v\rangle} \in \mathfrak{M}$, any propositional formula $\varphi\in \mathcal{L}_0$,  and any possible worlds $w, w'\in W$ satisfying $\varphi$, it holds:
\begin{itemize}
\item[] \DP{1a} if $w\leq_{\star \varphi} w'$($w<_{\star \varphi} w'$) then, for any piece of information $\xi \in \mathcal{L}_\leq(\star)$ s.t. $D,w \vDash [\star \varphi] \xi$, there is some world $w''\in \llbracket \varphi \rrbracket$ s.t. $D,w''\vDash [\star\varphi] \xi$ and $w''\leq w'$($w''< w'$);
\item[] \DP{1b} if $w\leq w'$ ($w< w'$ ) then, for any piece of information $\xi\in \mathcal{L}_\leq(\star)$ s.t. ${D,w \vDash [\star \varphi]\xi}$, there is some world $w''\in \llbracket \varphi \rrbracket$ s.t. $D,w''\vDash [\star\varphi] \xi$, and $w''\leq_{\star\varphi} w'$ ($w''<_{\star\varphi} w'$)
\end{itemize}
where $D = \langle M, \star\rangle$. If $\mathfrak{M}= \mathit{Mod}(\mathcal{L}_\leq)$, we say $\star$ is DP1-compliant.
\end{defn}

Definition~\ref{def:dp1plus} states that no information contained in the worlds satisfying $\varphi$ is lost due to a belief change regarding $\varphi$, similar to what \DP{1} tries to encode for Grove models. 
With this generalisation, we can characterise DP1-compliance in DPL.

\begin{prop}\label{prop:dp1plus}
Let $\mathfrak{D} = \langle \mathfrak{M}, \star\rangle$ be a class of dynamic models. The axiomatisation presented in Proposition~\ref{prop:CR1} is valid in $\mathfrak{D}$ iff $\star$ is $\mathfrak{M}$-DP1-compliant.
\end{prop}

\textcolor{blue}{
In the light of Propositions~\ref{prop:dp1syn} and \ref{prop:dp1plus}, it is easy to see that  the notion of $\mathfrak{M}$-DP1-compliance is, in fact, an adequate generalization of \DP{1} to  preference models, as observed in the Fact below.
\begin{fact}\label{fact:dp1syn}
Let $\mathfrak{M}\subseteq \mathit{Mod}(\mathcal{L}_\leq)$ be a class of preference models, $\star: \mathit{Mod}(\mathcal{L}_\leq)\times \mathcal{L}_0 \rightarrow \mathit{Mod}(\mathcal{L}_\leq)$ be a $\mathfrak{M}$-DP1-compliant dynamic operator, and $\mathfrak{D} = \langle \mathfrak{M}, \star\rangle$. For any propositional formulas $\varphi, \psi \in \mathcal{L}_0$ s.t. $\mathfrak{D}\vDash \varphi \rightarrow \psi$ and any dynamic preference formula $\xi \in \mathcal{L}_\leq(\star)$, it holds that
$$\mathfrak{D}\vDash [\star \psi]B(\xi~|~\varphi) \leftrightarrow B([\star \psi] \xi ~|~\varphi)$$
\end{fact}
}

\textcolor{blue}{Notice that for models in which every world has a characteristic formula, Definition~\ref{def:dp1plus} implies that the dynamic operator $\star$ satisfies \DP{1}. In other words, if we consider only Grove models and operators closed over Grove models, \DP{1a} and  \DP{1b} taken together are equivalent to \DP{1}.}

\textcolor{blue}{\begin{fact}\label{cor:CR1}
Let $\mathfrak{M}$ be a class of preference models s.t. for any $M\in\mathfrak{M}$ and any possible world $w$ in $M$ there is a characteristic formula $\xi_w\in \mathcal{L}_\leq(P)$, s.t. $M,w'\vDash \xi_w$ iff $w'=w$ and let $\star:Mod(\mathcal{L}_\leq)\times \mathcal{L}_0 \rightarrow Mod(\mathcal{L}_\leq)$ be a dynamic operator closed over $\mathfrak{M}$. It holds that $\star$ is $\mathfrak{M}$-DP1-compliant iff for any  $M\in \mathfrak{M}$, any propositional formula $\varphi$ and worlds $w,w'\in \llbracket\varphi\rrbracket$ it holds that $w\leq w'$ iff $w \leq_{\star \varphi} w'$.\end{fact}}

\paragraph{\textcolor{blue}{Postulate DP2}}
Similar results can be achieved for the other  postulates, i.e., we can provide characterisations of the other postulates by means of DPL axioms. Next, we provide characterisation for DP2.

\begin{defn}\label{def:dp2plus}\sloppy
Let $\mathfrak{M} \subseteq \mathit{Mod}(\mathcal{L}_\leq)$ be a class of preference models, and let $\star: \mathit{Mod}(\mathcal{L}_\leq)\times \mathcal{L}_0 \rightarrow \mathit{Mod}(\mathcal{L}_\leq)$ be a dynamic operator. We say $\star$ is DP2-compliant in $\mathfrak{M}$, or $\mathfrak{M}$-DP2-compliant, if, for any preference model $M = \langle W, \leq, v\rangle \in \mathfrak{M}$, any propositional formula $\varphi\in \mathcal{L}_0$  and any  possible worlds $w, w'\in W$ not satisfying $\varphi$, it holds:
\begin{itemize}
\item[] \DP{2a} if $w\leq_{\star \varphi} w'$ ($w<_{\star \varphi} w'$) then for any piece of information $\xi \in \mathcal{L}_\leq(\star)$ s.t. ${{D,w \vDash [\star \varphi] \xi}}$ there is some world $w''\not\in \llbracket \varphi \rrbracket$ s.t. $D,w''\vDash [\star\varphi] \xi$ and $w''\leq w'$ ($w''< w'$);
\item[] \DP{2b} if $w\leq w'$ ($w< w'$) then for any piece of information $\xi\in \mathcal{L}_\leq(\star)$ s.t. {${D,w \vDash [\star \varphi] \xi}$} there is some world $w''\not\in \llbracket \varphi \rrbracket$ s.t. {$D,w''\vDash [\star\varphi] \xi$} and $w\leq_{\star\varphi} w''$ ($w <_{\star\varphi} w''$)
\end{itemize}
where $D = \langle M, \star\rangle$. If $\mathfrak{M}= \mathit{Mod}(\mathcal{L}_\leq)$, we say $\star$ is DP2-compliant.
\end{defn}

As before, Definition~\ref{def:dp2plus} states that no information contained in the worlds satisfying $\neg \varphi$ is lost due to a belief change regarding $\varphi$, similar to what \DP{2} encodes for Grove models. It is easy to see that for models without propositionally indiscernible worlds, this condition is equivalent to \DP{2}, similar to proven in Fact~\ref{cor:CR1}. More yet, we can characterise the DP2-compliance using DPL.

\begin{prop}\label{prop:CR2}
Let $\mathfrak{D} = \langle \mathfrak{M}, \star\rangle$ be a class of dynamic models. The following axiom schemata is valid in $\mathfrak{D}$  for any $\varphi \in \mathcal{L}_0$ and $\xi \in \mathcal{L}_\leq(\star)$ iff $\star$ is $\mathfrak{M}$-DP2-compliant.
$$
\begin{array}{lcl}
{}[\star\varphi][\leq]\xi &\rightarrow& (\neg \varphi \rightarrow [\leq](\neg \varphi \rightarrow [\star\varphi] \xi))\\
{}[\star\varphi][<]\xi &\rightarrow& (\neg \varphi \rightarrow [<](\neg \varphi \rightarrow [\star\varphi] \xi))\\
{}[\leq][\star \varphi]\xi &\rightarrow& (\neg\varphi \rightarrow [\star\varphi][\leq](\neg \varphi \rightarrow \xi))\\
{}[<][\star \varphi]\xi &\rightarrow& (\neg\varphi \rightarrow [\star\varphi][<](\neg \varphi \rightarrow \xi))\\
\end{array}$$
\end{prop}

Notice that,  given the structural similarities between postulates \DP{1} and \DP{2},  the axioms presented in Proposition~\ref{prop:CR2} are similar to those presented in Proposition~\ref{prop:CR1} and also have similar interpretations.

\textcolor{blue}{Similar to Facts~\ref{cor:CR1} and \ref{fact:dp1syn}, it is easy to see that  our generalisation of \DP{2} implies \DP{2} for Grove-like models, namely those with characterising formulas, and preserves of incompatible conditional beliefs.}

\textcolor{blue}{\begin{fact}\label{fact:dp2syn}
Let $\mathfrak{M}\subseteq \mathit{Mod}(\mathcal{L}_\leq)$ be a class of preference models, $\star: \mathit{Mod}(\mathcal{L}_\leq)\times \mathcal{L}_0 \rightarrow \mathit{Mod}(\mathcal{L}_\leq)$ be a $\mathfrak{M}$-DP2-compliant dynamic operator, and $\mathfrak{D} = \langle \mathfrak{M}, \star\rangle$. For any propositional formulas $\varphi, \psi \in \mathcal{L}_0$ s.t. $\mathfrak{D}\vDash \varphi \rightarrow \neg \psi$ and any dynamic preference formula $\xi \in \mathcal{L}_\leq(\star)$, it holds that
$$\mathfrak{D}\vDash [\star \psi]B(\xi~|~\varphi) \leftrightarrow B([\star \psi] \xi ~|~\varphi)$$
\end{fact}}

\paragraph{\textcolor{blue}{Postulate DP3}}  Similarly, we can generalise the postulate \DP{3}. This postulate states that no $\neg \varphi$-world gets promoted after acquiring information that $\varphi$, or in other words, that all conditional belief consistent with $\varphi$ is maintained.

\begin{defn}\label{prop:dp3plus}
Let $\mathfrak{M}\subseteq \mathit{Mod}(\mathcal{L}_\leq)$ be a class of preference models,  and let $\star: \mathit{Mod}(\mathcal{L}_\leq)\times \mathcal{L}_0 \rightarrow \mathit{Mod}(\mathcal{L}_\leq)$ be a dynamic operator. We say $\star$ is DP3-compliant in $\mathfrak{M}$, or $\mathfrak{M}$-DP3-compliant, if for any preference model $M = \langle W, \leq, v\rangle \in \mathfrak{M}$, any propositional formula $\varphi\in \mathcal{L}_0$,  and any  possible worlds $w,w'\in W$, it holds that:
\begin{itemize}
\item[] \DP{3a} if $w \in \llbracket \varphi \rrbracket$, $w'\not\in \llbracket \varphi \rrbracket$ and $w < w'$, then for any information $\xi \in \mathcal{L}_\leq(\star)$ s.t. $D,w\vDash [\star \varphi]\xi$ there is some world $w''\in \llbracket \varphi \rrbracket$ s.t. $D,w''\vDash [\star\varphi] \xi$ and $w'' <_{\star \varphi} w'$.
\end{itemize}
where $D = \langle M, \star\rangle$. If $\mathfrak{M}= \mathit{Mod}(\mathcal{L}_\leq)$, we say $\star$ is DP3-compliant.
\end{defn}

As before, we can characterise DP3-compliance using DPL.

\begin{prop}\label{prop:CR3}
Let $\mathfrak{D} = \langle \mathfrak{M}, \star\rangle$ be a class of dynamic models. The following axiom schema is valid in $\mathfrak{D}$  for any $\varphi \in \mathcal{L}_0$ and $\xi \in \mathcal{L}_\leq(\star)$ iff $\star$ is $\mathfrak{M}$-DP3-compliant.
\vspace{-0.2cm}
$$
\begin{array}{lcl}
{}[\star \varphi][<](\varphi \rightarrow \xi) & \rightarrow &\neg \varphi \rightarrow [<] [\star \varphi] (\varphi \rightarrow \xi)
\end{array}$$
\end{prop}

\textcolor{blue}{As for \DP{1} and \DP{2}, our generalisation of \DP{3} does encode the idea behind the original postulate. This property may be observed in Fact~\ref{fact:dp3syn}, an interpretation of Darwiche and Pearl's \cite{darwiche} syntactic form of \DP{3} into DPL.}

\textcolor{blue}{\begin{fact}\label{fact:dp3syn}
Let $\mathfrak{M}\subseteq \mathit{Mod}(\mathcal{L}_\leq)$ be a class of preference models, $\star: \mathit{Mod}(\mathcal{L}_\leq)\times \mathcal{L}_0 \rightarrow \mathit{Mod}(\mathcal{L}_\leq)$ be a $\mathfrak{M}$-DP3-compliant dynamic operator, and $\mathfrak{D} = \langle \mathfrak{M}, \star\rangle$. For any propositional formula $\varphi \in \mathcal{L}_0$ and dynamic preference formula $\xi \in \mathcal{L}_\leq(\star)$, it holds that
$$\mathfrak{D}\vDash B(\varphi~|~[\star \varphi]\xi) \rightarrow [\star \varphi]B(\varphi ~|~\xi)$$
\end{fact}}

\paragraph{\textcolor{blue}{Postulate DP4}}  Similarly,  we can generalise the postulate \DP{4} to the following condition:

\begin{defn}\label{def:dp4plus}
Let $\mathfrak{M}\subseteq \mathit{Mod}(\mathcal{L}_\leq)$ be a class of preference models,  and let $\star: \mathit{Mod}(\mathcal{L}_\leq)\times \mathcal{L}_0 \rightarrow \mathit{Mod}(\mathcal{L}_\leq)$ be a dynamic operator. We say $\star$ is DP4-compliant in $\mathfrak{M}$, or $\mathfrak{M}$-DP4-compliant, if for any preference model $M = \langle W, \leq, v\rangle \in \mathfrak{M}$, any propositional formula $\varphi\in \mathcal{L}_0$, and any  possible worlds $w,w'\in W$, it holds that:
\begin{itemize}
\item[] \DP{4a} if $w \in \llbracket \varphi \rrbracket$, $w'\not\in \llbracket \varphi \rrbracket$ and $w\leq w'$, then for any information $\xi \in \mathcal{L}_\leq(\star)$ s.t. $D,w\vDash [\star] \xi$ there is some world $w''\in \llbracket \varphi \rrbracket$ s.t. $D,w''\vDash [\star\varphi] \xi$ and $w'' \leq_{\star \varphi} w'$.
\end{itemize}
where $D = \langle M, \star\rangle$. If $\mathfrak{M}= \mathit{Mod}(\mathcal{L}_\leq)$, we say $\star$ is DP4-compliant.
\end{defn}

Again, we can characterise DP4-compliance.

\begin{prop}\label{prop:CR4}
Let $\mathfrak{D} = \langle \mathfrak{M}, \star\rangle$ be a class of dynamic models. The following axiom schemata  is valid in $\mathfrak{D}$  for any $\varphi \in \mathcal{L}_0$ and $\xi \in \mathcal{L}_\leq(\star)$ iff $\star$ is $\mathfrak{M}$-DP4-compliant.
\vspace{-0.2cm}
$$
\begin{array}{lcl}
{}[\star \varphi][\leq](\varphi \rightarrow \xi) & \rightarrow &\neg \varphi \rightarrow [\leq] [\star \varphi] (\varphi\rightarrow \xi)
\end{array}$$
\end{prop}

\textcolor{blue}{As before, our generalisation of \DP{4} also encodes the idea behind the original postulate.}

\textcolor{blue}{\begin{fact}\label{fact:dp4syn}
Let $\mathfrak{M}\subseteq \mathit{Mod}(\mathcal{L}_\leq)$ be a class of preference models, $\star: \mathit{Mod}(\mathcal{L}_\leq)\times \mathcal{L}_0 \rightarrow \mathit{Mod}(\mathcal{L}_\leq)$ be a $\mathfrak{M}$-DP4-compliant dynamic operator, and $\mathfrak{D} = \langle \mathfrak{M}, \star\rangle$. For any propositional formula $\varphi \in \mathcal{L}_0$ and dynamic preference formula $\xi \in \mathcal{L}_\leq(\star)$, it holds that
$$\mathfrak{D}\vDash \neg B(\neg \varphi~|~[\star \varphi]\xi) \rightarrow \neg [\star \varphi]B(\neg \varphi ~|~\xi)$$
\end{fact}} 

\paragraph{\textcolor{blue}{Postulate REC}}  Below we provide a characterisation in DPL for Nayak et al.'s \cite{nayak2003dynamic} Recalcitrance, or \REC{}.

\begin{defn}\label{def:recplus}
Let $\mathfrak{M}\subseteq \mathit{Mod}(\mathcal{L}_\leq)$ be a class of preference models,  and let $\star: \mathit{Mod}(\mathcal{L}_\leq)\times \mathcal{L}_0 \rightarrow \mathit{Mod}(\mathcal{L}_\leq)$ be a dynamic operator. We say $\star$ is $\mathfrak{M}$-Rec-compliant, if, for any preference model $M = \langle W, \leq, v\rangle \in \mathfrak{M}$, any propositional formula $\varphi\in \mathcal{L}_0$,  and any  possible worlds $w,w'\in W$, it holds that:
\begin{itemize}
\item[] \REC{'} if $w \in \llbracket \varphi \rrbracket$ and $w' \not\in \llbracket \varphi \rrbracket$, then $w' \not\leq_{\star \varphi} w$ and for any information $\xi \in \mathcal{L}_\leq$ s.t. $D,w\vDash [\star \varphi]\xi$ there is some world $w''\in \llbracket \varphi\rrbracket$ s.t. $D,w''\vDash [\star \varphi] \xi$ and $w'' \leq_{\star \varphi} w'$.
\end{itemize}
where $D = \langle M, \star\rangle$.  If $\mathfrak{M}= \mathit{Mod}(\mathcal{L}_\leq)$, we say $\star$ is Rec-compliant.
\end{defn}

From this encoding, we obtain the following characterisation.

\begin{prop}\label{prop:Rec}
Let $\mathfrak{D} = \langle \mathfrak{M}, \star\rangle$ be a class of dynamic models. The following axiom schemata  is valid in $\mathfrak{D}$  for any $\varphi \in \mathcal{L}_0$ and $\xi \in \mathcal{L}_\leq(\star)$ iff $\star$ is $\mathfrak{M}$-Rec-compliant.
\vspace{-0.2cm}
$$
\begin{array}{lcl}
{}[\star\varphi][<]\xi &\rightarrow& (\neg \varphi \rightarrow A(\varphi \rightarrow [\star\varphi] \xi))\\
{}\varphi &\rightarrow &([\star\varphi][\leq]\varphi)\\
\end{array}$$
\end{prop}

\textcolor{blue}{Similar to Nayak et al.~\cite{nayak2003dynamic}'s \REC{}, our \REC{'} postulate guarantees the maximal preservation of an adopted belief $\psi$ in all conditions that do not contradict it.}

\textcolor{blue}{\begin{fact}\label{fact:recsyn}
Let $\mathfrak{M}\subseteq \mathit{Mod}(\mathcal{L}_\leq)$ be a class of preference models, $\star: \mathit{Mod}(\mathcal{L}_\leq)\times \mathcal{L}_0 \rightarrow \mathit{Mod}(\mathcal{L}_\leq)$ be a $\mathfrak{M}$-Rec-compliant dynamic operator, and $\mathfrak{D} = \langle \mathfrak{M}, \star\rangle$. For any propositional formula $\varphi \in \mathcal{L}_0$, and dynamic preference formula $\xi \in \mathcal{L}_\leq(\star)$, it holds that
$$\mathfrak{D}\vDash E(\varphi \wedge [\star\varphi] \xi) \rightarrow [\star \varphi]B( \varphi~|~\xi)$$
\end{fact}} 

%
\subsection{Postulates of Iterated Belief Contraction in DPL}
\label{subsec:ibc-in-DPL}
%

Regarding the postulates for iterated belief contraction, 
we present DPL characterisations for Chopra et al.'s \cite{chopra2008iterated} \CR{1} - \CR{4}, and Ramachandran et al.'s \cite{ramachandran2012three} \LC{}. 
Let us begin with the representation of Grove's \cite{grove} characterisation of contraction in DPL, as presented in Section~\ref{sec:ibr}.

\begin{prop}\label{prop:GR}
Let $\mathfrak{C}$ be a class of dynamic operators satisfying $\mathtt{(GR)}$ and $\mathfrak{M}\subseteq \mathit{Mod}(\mathcal{L}_\leq)$ be a class of preference models. The following axiom schema is valid in $\langle \mathfrak{M}, \mathfrak{C}_{\mathtt{(GR)}}\rangle$, for any propositional formula $\varphi \in \mathcal{L}_0$ and formula $\xi \in \mathcal{L}_\leq(\star)$.
\vspace{-0.2cm}
$$
\begin{array}{lcl}
{}(\mu\neg \varphi \vee \mu \top) &\leftrightarrow & [\star\varphi]\mu\top
\end{array}$$
\end{prop}

Notice that Chopra et al.'s \cite{chopra2008iterated} \CR{1} - \CR{4} are variants of Darwiche and Pearl's \cite{darwiche}'s \DP{1}-\DP{4}. In fact, \CR{1} is the same postulate as \DP{2} and \CR{2} is the same as \DP{1}, while \CR{3} and \DP{3}, and \CR{4} and \DP{4} are structurally similar. In fact, we know that \CR{3} and \CR{4} are dual forms of \DP{3} and \DP{4}, respectively \cite{booth2019iterated}. Since DP1- and DP2-compliance have already been defined, we will only focus on \CR{3} and \CR{4}. \textcolor{blue}{As for \DP{4} and \DP{4}, our generalisation of postulates \CR{3} and \CR{4} also satisfy the original syntactic postulates of Chopra et al.'s \cite{chopra2008iterated}. Since these results are very similar to Fact~\ref{fact:dp3syn} and Fact~\ref{fact:dp4syn}, we omit the results about the suitability of our generalised postulates.}

\paragraph{\textcolor{blue}{Contraction Postulate CR3}} We can define the notion of CR3-compliance of a dynamic operator in regards to a class of preference models based on the notion of DP3-compliance.

\begin{defn}\label{def:cr3plus}
Let $\mathfrak{M}\subseteq \mathit{Mod}(\mathcal{L}_\leq)$ be a class of preference models,  and let $\star: \mathit{Mod}(\mathcal{L}_\leq)\times \mathcal{L}_0 \rightarrow \mathit{Mod}(\mathcal{L}_\leq)$ be a dynamic operator. We say $\star$ is $\mathfrak{M}$-CR3-compliant, if,  for any preference model $M = \langle W, \leq, v\rangle \in \mathfrak{M}$, any propositional formula $\varphi\in \mathcal{L}_0$,   and any  possible worlds $w,w'\in W$, it holds that:
\begin{itemize}
\item[] \CR{3a} if $w \not\in \llbracket \varphi \rrbracket$, $w'\in \llbracket \varphi \rrbracket$ and $w < w'$, then for any information $\xi \in \mathcal{L}_\leq$ s.t. $D,w\vDash [\star \varphi] \xi$ there is some world $w''\not\in \llbracket \varphi \rrbracket$ s.t. $D,w''\vDash [\star\varphi] \xi$ and $w'' <_{\star \varphi} w'$
\end{itemize}
where $D = \langle M, \star\rangle$. If $\mathfrak{M}= \mathit{Mod}(\mathcal{L}_\leq)$, we say $\star$ is CR3-compliant.
\end{defn}

Hence, similarly to DP3-compliance, we can characterise CR3-compliance.

\begin{prop}\label{prop:cont3}
Let $\mathfrak{D} = \langle \mathfrak{M}, \star\rangle$ be a class of dynamic models. The following axiom schemata  is valid in $\mathfrak{D}$  for any $\varphi \in \mathcal{L}_0$ and $\xi \in \mathcal{L}_\leq(\star)$ iff $\star$ is $\mathfrak{M}$-CR3-compliant.
\vspace{-0.2cm}
$$
\begin{array}{lcl}
{}[\star \varphi][<](\neg \varphi \rightarrow \xi) & \rightarrow &\varphi \rightarrow [<] [\star \varphi] (\neg \varphi \rightarrow \xi)
\end{array}$$
\end{prop}

\paragraph{\textcolor{blue}{Contraction Postulate CR4}} As \CR{4} is similar variation of \DP{4}, it is easy to define the notion of CR4-compliance.

\begin{defn}\label{def:cr4plus}
Let $\mathfrak{M}\subseteq \mathit{Mod}(\mathcal{L}_\leq)$ be a class of preference models,  and let $\star: \mathit{Mod}(\mathcal{L}_\leq)\times \mathcal{L}_0 \rightarrow \mathit{Mod}(\mathcal{L}_\leq)$ be a dynamic operator. We say $\star$ is $\mathfrak{M}$-CR4-compliant, if, for any preference model $M = \langle W, \leq, v\rangle \in \mathfrak{M}$, any propositional formula $\varphi\in \mathcal{L}_0$,  and any  possible worlds $w,w'\in W$, it holds that:
\begin{itemize}
\item[] \CR{4a} if $w \not\in \llbracket \varphi \rrbracket$, $w'\in \llbracket \varphi \rrbracket$ and $w\leq w'$, then for any information $\xi \in \mathcal{L}_\leq$ s.t. $D,w\vDash [\star \varphi] \xi$ there is some world $w''\not\in \llbracket \varphi \rrbracket$ s.t. $D,w''\vDash [\star\varphi] \xi$ and $w'' \leq_{\star \varphi} w'$.
\end{itemize}
where $D = \langle M, \star\rangle$. If $\mathfrak{M}= \mathit{Mod}(\mathcal{L}_\leq)$, we say $\star$ is CR4-compliant.
\end{defn}

Again, it follows that for models without propositionally indiscernible worlds, this condition is equivalent to \DP{4}.

\begin{prop}\label{prop:cont4}
Let $\mathfrak{D} = \langle \mathfrak{M}, \star\rangle$ be a class of dynamic models. The following axiom schemata  is valid in $\mathfrak{D}$  for any $\varphi \in \mathcal{L}_0$ and $\xi \in \mathcal{L}_\leq(\star)$ iff $\star$ is $\mathfrak{M}$-CR4-compliant.
\vspace{-0.2cm}
$$
\begin{array}{lcl}
{}[\star \varphi][\leq](\neg \varphi \rightarrow \xi) & \rightarrow & \varphi \rightarrow [\leq] [\star \varphi] (\neg \varphi\rightarrow \xi)
\end{array}$$
\end{prop}

\paragraph{\textcolor{blue}{Contraction Postulate LC}} Finally, we can represent Lexicographic Contraction in DPL. Notice that \LC{} states how the preference relation must be changes in terms of the maximal chains of worlds in the model that either satisfy $\varphi$ or satisfy $\neg \varphi$. As such, we define the notion of LC-compliance. \textcolor{blue}{In Definition~\ref{def:lcplus} notice that postulate \LC{'} is the same postulate proposed in Section~\ref{sec:dyn} with a slight difference in presentation to account for the notion of satisfaction of a formula in a dynamic model $D$.} 

\begin{defn}\label{def:lcplus}
Let $\mathfrak{M}\subseteq \mathit{Mod}(\mathcal{L}_\leq)$ be a class of preference models,  and let $\star: \mathit{Mod}(\mathcal{L}_\leq)\times \mathcal{L}_0 \rightarrow \mathit{Mod}(\mathcal{L}_\leq)$ be a dynamic operator. We say $\star$ is $\mathfrak{M}$-LC-compliant iff for any preference model $M = \langle W, \leq, v\rangle \in \mathfrak{M}$, any propositional formula $\varphi\in \mathcal{L}_0$  and any  possible worlds $w,w'\in W$, it holds that:
\begin{itemize}
\item[] \LC{'} Let $\xi,\chi $ be members of $\{\varphi, \neg\varphi\}$ - not necessarily distinct. If $D,w\vDash \xi$ and $D,w' \vDash \chi$, then $w \leq_{\star \varphi} w'$ iff the maximal length of a chain of worlds in $\llbracket\xi\rrbracket$ which ends in $w$ is smaller or equal than to the maximal length of a chain of worlds in $\llbracket\chi\rrbracket$ which ends in $w'$.
\end{itemize}
where $D = \langle M, \star\rangle$. If $\mathfrak{M}= \mathit{Mod}(\mathcal{L}_\leq)$, we say $\star$ is LC-compliant.
\end{defn}

We can, then, provide axioms to encode LC-compliance in DPL. Notice that the axioms below simply state that the result of an LC-compliant operator on a model order the worlds based on the length of the maximal $\varphi$ and $\neg \varphi$ chains in the model\footnote{Remember that $M,w\vDash \mu dg_\xi(i)$ iff there is a maximal chain of $\xi$-worlds ending in $w$, according to Lemma~\ref{lem:dg}.}, as stated in postulate \LC{'}.

\begin{prop}\label{prop:LC}
Let $\mathfrak{D} = \langle \mathfrak{M}, \star\rangle$ be a class of dynamic models. The following axiom schemata is valid in $\mathfrak{D}$, for all $n \in \mathbb{N}$, $\varphi \in \mathcal{L}_0$, and $\xi \in\mathcal{L}_\leq(\star)$  if $\star$ is $\mathfrak{M}$-LC-compliant.
\vspace{-0.2cm}
$$
\begin{array}{lcll}

{} [\star \varphi][\leq] \xi &\rightarrow &\displaystyle\bigwedge_{i=1}^{n}\bigwedge_{j=i}^{n} \mu dg_{\varphi}(j) \rightarrow A (\mu dg_{\varphi}(i)\rightarrow  [\star \varphi]\xi) \wedge\\
&&\displaystyle\bigwedge_{i=1}^{n}\bigwedge_{j=i}^{n} \mu dg_{\neg \varphi}(j) \rightarrow A (\mu dg_{\neg \varphi}(i)\rightarrow  [\star \varphi]\xi) \wedge\\

&&\displaystyle\bigwedge_{i=1}^{n}\bigwedge_{j=i}^{n} \mu dg_\varphi(j) \rightarrow A (\mu dg_{\neg \varphi}(i)\rightarrow  [\star \varphi]\xi) \wedge\\
&&\displaystyle\bigwedge_{i=1}^{n} \bigwedge_{j=i}^{n} \mu dg_{\neg \varphi(j)} \rightarrow A (\mu dg_{\varphi}(i)\rightarrow  [\star \varphi]\xi)\\

{} [\star \varphi][<] \xi &\rightarrow &\displaystyle\bigwedge_{i=1}^{n}\bigwedge_{j=i+1}^{n} \mu dg_{\varphi}(j) \rightarrow A (\mu dg_{\varphi}(i)\rightarrow  [\star \varphi]\xi) \wedge\\
&&\displaystyle\bigwedge_{i=1}^{n}\bigwedge_{j=i+1}^{n} \mu dg_{\neg \varphi}(j) \rightarrow A (\mu dg_{\neg \varphi}(i)\rightarrow  [\star \varphi]\xi) \wedge\\
&&\displaystyle\bigwedge_{i=1}^{n}\bigwedge_{j=i+1}^{n} \mu dg_\varphi(j) \rightarrow A (\mu dg_{\neg \varphi}(i)\rightarrow  [\star \varphi]\xi) \wedge\\
&&\displaystyle\bigwedge_{i=1}^{n} \bigwedge_{j=i+1}^{n} \mu dg_{\neg \varphi}(j) \rightarrow A (\mu dg_{\varphi}(i)\rightarrow  [\star \varphi]\xi)\\

{} [\star \varphi][\leq] \xi &\leftarrow & (\mu dg_{\neg\varphi}(n) \vee \mu dg_\varphi(n))\wedge \displaystyle\bigwedge_{i=1}^{n} A (\mu dg_{\varphi}(i)\rightarrow  [\star \varphi]\xi) \wedge \\
&&\displaystyle\bigwedge_{i=1}^{n} A (\mu dg_{\neg \varphi}(i)\rightarrow  [\star \varphi]\xi)\\

{} [\star \varphi][<] \xi &\leftarrow & (\mu dg_{\neg\varphi}(n) \vee \mu dg_\varphi(n))\wedge\displaystyle\bigwedge_{i=1}^{n-1} A (\mu dg_{\varphi}(i)\rightarrow  [\star \varphi]\xi) \wedge\\
&&\displaystyle\bigwedge_{i=1}^{n-1} A (\mu dg_{\neg \varphi}(i)\rightarrow  [\star \varphi]\xi)\\

\end{array}$$
\end{prop}

\textcolor{blue}{For guidance in comparing our generalised postulates to the original ones proposed for  Grove's systems of spheres, we list all postulates discussed in this work with the corresponding generalization in Table~\ref{tab:post}.
\begin{table}[!ht]
\centering
\caption{Original postulates and corresponding postulates for preference models.}
\label{tab:post}
\begin{tabular}{lcl} \\ \hline
{\small{Postulate}} & {\small{Original work}} & {\small{Corresponding postulates}}\\ \hline \noalign{\vskip 2mm}
\FAITH{} & \cite{grove} & --\\
\DP{1} & \cite{darwiche} & \DP{1a} and \DP{1b} (Definition~\ref{def:dp1plus})\\
\DP{2} & \cite{darwiche} & \DP{2a} and \DP{2b} (Definition~\ref{def:dp2plus})\\
\DP{3} & \cite{darwiche} & \DP{3a} (Definition~\ref{prop:dp3plus})\\
\DP{4} & \cite{darwiche} & \DP{4a} (Definition~\ref{def:dp4plus})\\
\REC{} & \cite{nayak2003dynamic} & \REC{'} (Definition~\ref{def:recplus})\\
\GR{} & \cite{grove} & --\\
\CR{1} & \cite{chopra2008iterated} & \DP{2a} and \DP{2b} (Definition~\ref{def:dp2plus})\\
\CR{2} & \cite{chopra2008iterated} & \DP{1a} and \DP{1b} (Definition~\ref{def:dp1plus})\\
\CR{3} & \cite{chopra2008iterated} & \CR{3a} (Definition~\ref{def:cr3plus})\\
\CR{4} & \cite{chopra2008iterated} & \CR{4a} (Definition~\ref{def:cr4plus})\\
\LC{} & \cite{ramachandran2012three} & \LC{'} (Definition~\ref{def:lcplus})\\
\end{tabular}
\end{table}}


\subsection{Deriving axiomatisations for DPL by Iterated Belief Change Postulates}\label{sub:der}

In the following, we investigate how we can derive sound axiomatisations for the logic $\mathcal{L}_\leq(\star)$, given the postulates satisfied by $\star$. First, let us properly define the notion of a logic induced by classes of dynamic preference models and by sets of axioms as proposed earlier.
\begin{defn}
Let $\mathfrak{D} = \langle \mathfrak{M},\mathfrak{C}\rangle$ be a class of dynamic models \textcolor{blue}{and $\star$ a symbol for dynamic operators, we call the logic of $\mathfrak{D}$}, or the logic defined by $\mathfrak{D}$, as the set $$Log(\mathfrak{D}) =  \{\varphi \in \mathcal{L}_\leq(\star) ~|~ \mathfrak{D}\vDash \varphi\}$$
\end{defn}
It is easy to see that the logic defined by a class of dynamic operators can be constructed from the logics induced by each operator individually. More generally, we have the following.

\begin{prop}\label{prop:intersect}
Let $\mathfrak{M}\subseteq \mathit{Mod}(\mathcal{L}_\leq)$ be a class of preference models and $\mathcal{C} = \{\mathfrak{C}_i ~|~i\in I\}$ a family of classes of dynamic operators indexed by some set $I$, which are closed over $\mathfrak{M}$. \textcolor{blue}{$$Log(\langle \mathfrak{M}, \bigcup_{i\in I} \mathfrak{C}_i ) = \bigcap_{i \in I} Log(\langle \mathfrak{M},\mathfrak{C}_i\rangle)$$} 
\end{prop}
Proposition~\ref{prop:intersect} implies that the logic defined by a class of dynamic models $\mathfrak{D} = \langle \mathfrak{M}, \mathfrak{C}\rangle$ is completely determined by each individual dynamic operator $\star \in \mathfrak{C}$. As such, if we wish to study the properties of, say, DP1-compliant operators over a class of preference models $\mathfrak{M}$ using DPL, it suffices to study the properties of each individual DP1-compliant operator over this class of models. Now, we will concern ourselves with the logic defined by a set of axioms - or a proof system. This will allow us to investigate the properties of soundness and completeness for DPL.

\begin{defn}\label{def:logA}
Let $A \subset \mathcal{L}_\leq(\star)$ be a set of formulas (or axioms). We define the logic of $A$, or defined by the system $A$, denoted $Log(A)$, as the smaller set of formulas containing $A$ that is closed by \textit{modus ponens} and necessitation rules, i.e., (i) if $\varphi, \varphi\rightarrow \psi \in Log(A)$ then $\psi \in Log(A)$ and (ii) if $\psi \in Log(A)$ then $\Box\varphi\in Log(A)$, with $\Box \in \{A,[\leq],[<],[\star \varphi]\mbox{ for any }\varphi\in \mathcal{L}_0\}$. We call $A$ an axiom system for $Log(A)$.
\end{defn}

With that, we can define the notions of an axiomatisation, or of a proof system, being sound or complete, as usual.

\begin{defn}
Let $A \subset \mathcal{L}_\leq(\star)$ be a set of formulas and $\mathfrak{D}$ a class of dynamic models. We say that
\begin{itemize} 
\item $A$ is sound in regards to $\mathfrak{D}$, if $Log(A) \subseteq Log(\mathfrak{D})$
\item $A$ is complete in regards to $\mathfrak{D}$, if $Log(\mathfrak{D}) \subseteq Log(A)$
\end{itemize}
\end{defn}

It is easy to see that the union of sound axiomatisations for a set of classes of dynamic models results in a sound axiomatisation for the intersection of such classes.

\begin{theo}\label{teo:correct}
Let $\mathfrak{M}$ be a class of preference models, $\mathcal{C} = \{\mathfrak{C}_i ~|~i\in I\}$ be a family of classes of dynamic operators, which are closed over $\mathfrak{M}$, and $\mathcal{A} = \{A_i ~|~ i\in I\}$ a family of sound axiom systems for $\mathcal{C}$, i.e., $Log(A_i)\subseteq Log(\langle \mathfrak{M}, \mathfrak{C}_i\rangle)$, both indexed by some set $I$. \textcolor{blue}{$$Log(\bigcup_{i\in I} A_i) \subseteq Log(\langle \mathfrak{M}, \bigcap_{i\in I}\mathfrak{C}_i)$$}
\end{theo}

In Theorem~\ref{teo:correct}, we prove that we can combine the axiomatisations that represent each postulate (presented in Propositions~\ref{prop:BasicAxiom}, \ref{prop:CR1}, \ref{prop:CR2} \ref{prop:CR3} and so on) into a single axiomatisation that is sound to the class of operators that satisfy all postulates at once. As such, we can use the results in this section to obtain sound logics for classes of dynamic operators, such as the logic defined by Darwich and Pearl's Iterated Belief Revision \cite{darwiche} operators or Nayak et al.'s Lexicographic Revision operators \cite{nayak2003dynamic}.

Notice that, while we were able to characterise Iterated Belief Change postulates in DPL,  in Propositions~\ref{prop:dp1plus}, \ref{prop:CR2}, \ref{prop:CR3}, \ref{prop:CR4}, \ref{prop:Rec}, \ref{prop:GR}, \ref{prop:cont3} and \ref{prop:cont4}, 
this does not guarantee that the resulting logic is complete. In fact, it is not easy to obtain a completeness result for the axiomatisations since, given the semantics of DPL based on dynamic models, one such proof would require the construction of a dynamic model serving as counter-example for any non-theorem of the logic. While the technique of filtrated canonical models, as used by Souza to prove completeness for Preference Logic \cite{souzaphd}, does indicate ways to the construction of such model, it is not clear how to translate the obtained filtrated canonical model into a dynamic model.

More yet, even if we obtain complete axiomatisations for some postulates, it is not clear whether we can obtain a general completeness result for the conjoined axiomatisations, such as done in Theorem~\ref{teo:correct} for soundness. 

In the light of Theorem~\ref{teo:correct}, we can apply the results obtained in Propositions~\ref{prop:BasicAxiom}, \ref{prop:CR1},~\ref{prop:CR2} and ~\ref{prop:Rec} to obtain an axiomatisation to DPL of Lexicographic Revision.

\begin{corolary}\label{cor:axiomRU}
Preference Logic extended with lexicographic revision is soundly axiomatised by the axiomatization of $\mathcal{L}_\leq$ extended by the axioms and rules below.
$$\begin{array}{lcl}
{}[\Uparrow\varphi]p &\leftrightarrow& p\\
{}[\Uparrow\varphi](\psi \wedge \xi)&\leftrightarrow& [\Uparrow\varphi]\psi \wedge [\Uparrow\varphi]\xi\\
{}[\Uparrow\varphi]\neg \xi&\leftrightarrow&\neg [\Uparrow\varphi]\xi\\
{}[\Uparrow\varphi]A \xi &\leftrightarrow& A [\Uparrow\varphi] \xi\\
{}[\star\varphi][\leq]\xi &\rightarrow& (\varphi \rightarrow [\leq](\varphi \rightarrow [\star\varphi] \xi))\\
{}[\star\varphi][<]\xi &\rightarrow& (\varphi \rightarrow [<](\varphi \rightarrow [\star\varphi] \xi))\\
{}[\leq][\star\varphi]\xi &\rightarrow& (\varphi \rightarrow [\star\varphi][\leq](\varphi \rightarrow \xi))\\
{}[<][\star\varphi]\xi &\rightarrow& (\varphi \rightarrow [\star\varphi][<](\varphi \rightarrow \xi))\\
{}[\star\varphi][\leq]\xi &\rightarrow& (\neg \varphi \rightarrow [\leq](\neg \varphi \rightarrow [\star\varphi] \xi))\\
{}[\star\varphi][<]\xi &\rightarrow& (\neg \varphi \rightarrow [<](\neg \varphi \rightarrow [\star\varphi] \xi))\\
{}[\leq][\star \varphi]\xi &\rightarrow& (\neg\varphi \rightarrow [\star\varphi][\leq](\neg \varphi \rightarrow \xi))\\
{}[<][\star \varphi]\xi &\rightarrow& (\neg\varphi \rightarrow [\star\varphi][<](\neg \varphi \rightarrow \xi))\\
{}[\star\varphi][<]\xi &\rightarrow& (\neg \varphi \rightarrow A(\varphi \rightarrow [\star\varphi] \xi))\\
{}\varphi &\rightarrow &([\star\varphi][\leq]\varphi)\\
\vdash \xi & \Rightarrow & \vdash [\star \varphi]\xi
\end{array}$$
\end{corolary}
 It is not difficult to see that the complete axiomatisation presented in Figure~\ref{fig:axiomRU} at Section~\ref{sec:dyn} is a simplification of the axiomatisation of Corollary~\ref{cor:axiomRU} obtained using our method. The same stands for axiomatisations for Lexicographic Contraction (Proposition~\ref{prop:LC}), when compared to the one obtained by Souza \cite{souzaphd} for finite models with chains of a bounded size.

\section{Related Work}
\label{sec:rel}

To our knowledge, the work of Segerberg~\citeonline{segerberg} is the first to propose the integration of belief revision operations within an epistemic logic, with his proposal of Dynamic Doxastic Logic (DDL). This integration is important because it allows one to analyse the effects of introspection, and other related phenomena, in the logic of belief change. A famous example of such interaction is the analysis of Moore sentences in the logic of belief change, which shows that AGM's postulates are incompatible in the face of introspection \cite{lindstrom:rabinowicz}. In this work, Segerberg provides a set of axioms , which corresponds to encodings within his logic of AGM's postulates for belief change. Our work is linked to that approach by investigating these correspondences for dynamic belief change, based on iterated belief change postulates\textcolor{blue}{, instead of AGM belief change, as pursued by Segerberg}. 

In the context of DDL, Cantwell \cite{cantwell1999some} defines some iterated belief revision operators as change operations in hypertheories \cite{lindstrom:rabinowicz} and he shows how these operations can be axiomatically characterised in DDL. Our work differs from his in that we analyse how some well known-postulates can be characterised in our logic and not how to encode specific constructions. Our logic also has the advantage of being more expressive because it can encode some notion of degrees of belief \cite{souzaphd}, which cannot be expressed in DDL.

\textcolor{blue}{Another trend of research on using modal logics to study (and axiomatise) iterated belief change operators or policies is the work on Dynamic Epistemic Logics.} Inspired by Rott \cite{rott:shift}, Van Benthem \cite{van2007dynamic} proposed the codification of some iterated belief revision operators within a Dynamic Epistemic Logic (DEL). 
This work was further extended by Girard \cite{girard2008modal}, Liu \cite{liu2011reasoning}, and Souza et al.~\cite{souzakr} who studied the use of DPL to encode several (relational) belief revision policies. Similarly, Baltag and Smets \cite{BAL08} used a logic similar to DPL to encode different notions for knowledge and belief, based on Board's work \cite{board2004dynamic}. These authors show how different iterated belief revision operators can be simulated using DEL action models and product update. 

Further, 
Girard and Rott \cite{girard2014belief} propose a DPL for studying belief revision.  
The authors encode several iterated belief revision policies using General Dynamic Dynamic Logic \cite{girard2012general} 
and show that reduction axioms can be obtained for them in the same fashion as \cite{pref}.

In the related literature, all these works following the DEL tradition define operations semantically in their logic and either provide axiomatisations by means of crafting the axioms or by encoding these operations using a variation of dynamic logic to obtain reduction axioms. These works are informed by well-known results in the area of belief change to choose appropriate operations and then encode these operations in their logics.

On the other hand, our work investigates how DPL can be used to characterise properties of dynamic belief change operators\textcolor{blue}{, instead of applying the characterisation in Belief Change theory to construct a logic. To our knowledge, our work is the first to do so for dynamic belief change operators. While other work, such as that of Darwiche and Pearl's \cite{darwiche} and of Jin and Thielscher \cite{JIN06} and others, have investigated semantic characterisation of iterated belief change postulates, these characterisations have been pursued in an extra-logical framework, i.e., outside of the object language used to the  specify the agent's beliefs. As discussed before, these move to specifying belief change has important consequences on the expressiveness of the theory.}

\textcolor{blue}{Our work is also connected to that of Souza et al.~\cite{souza2019iterated,souza:dali19} that uses the connection between preference models and Liu's priority graphs~\cite{liu2011reasoning} to belief change postulates as structural properties on transformations in priority graphs. However, These authors show that transformations on priority graphs are a limited representation for changes of dynamic operators and some of the well-known postulates of the area cannot be encoded using them. How our generalisations can be connected to that work is still an open topic of research. It seems doubtful that we can establish significant connections between our axiomatic characterisations and the structural constraints  imposed by postulates in the realisation of dynamic operators as transformations in priority graphs.}

\textcolor{blue}{As far as we know, the only work that investigates the encoding of postulates from Belief Change as axioms of a logic is the work of Segerberg~\cite{segerberg,SEG01} on DDL. In \cite{SEG01}, the author axiomatizes the logic DDL and proves the completeness of the logic in regards to an SOS-based semantics, showing thus that their axioms are accurate representations of AGM's postulates within their logic. While Segerberg's work is very similar to ours in intent, the authors do not consider iterated belief change postulates and their representation in the logic - the main focus of our work. Notice that, while DDL could, in principle, be used as a foundation for our study, this logic is less expressive than DPL. Also, by choosing to employ DPL instead of DDL as a foundational logic, our work can be connected to the work representing different mental attitudes in this logic \cite{van2009everything,liu:deontics,souza2017dynamic} allowing the application of theoretically founded dynamic operators to dynamic phenomena for different mental attitudes.}

%% file: proofs.tex
\section{Proofs of selected results in the paper}\label{ap:proofs}

All proofs related to postulates \DP{2}, \DP{4}, \CR{3} and \CR{4} will be omitted.  The proofs of these results employ similar arguments used in the proofs for the analogous results for postulate \DP{1}, \DP{2}, and \DP{3}. 

In the proofs below, to simplify the argumentation, we will usually employ existential versions of the axioms discussed in Section~\ref{sec:iter}, obtained by the contraposition of the original axioms. The proof of Fact~\ref{fact:13pdl} below requires familiarity with Propositional Dynamic Logic (PDL).

\begin{reffact}{\ref{fact:13pdl}}
Lexicographic Contraction cannot be encoded by means of PDL programs
\end{reffact}
\begin{proof}
Let $M = \langle W, \leq, v\rangle$ be a preference model and $\varphi \in \mathcal{L}_0$ a propositional formula. The preference relation $\leq_{\Downarrow \varphi}$ resulting from the application of Lexicographic Contraction on $M$ can be computed as: 
$$\begin{array}{ll}
\leq_{\Downarrow \varphi} = & \bigcup \{\llbracket(?\mu dg_\varphi(i);\top;?\mu dg_{\varphi}(j))\rrbracket~| i \leq j\}^M \cup\\
 & \bigcup \{\llbracket(?\mu dg_{\neg \varphi(i)};\top;?\mu dg_{\neg \varphi}(j))\rrbracket~| i \leq j\}^M \cup\\
& \bigcup \{\llbracket(?\mu dg_\varphi(i);\top;?\mu dg_{\neg \varphi}(j))\rrbracket~| i \leq j\}^M \cup\\
&  \bigcup \{\llbracket(?\mu dg_{\neg \varphi(i)};\top;?\mu dg_{\varphi}(j))\rrbracket~|~ i\leq j\}^M
\end{array} $$

As PDL is a logic of regular programs and the construction above corresponds with the language $L = \{a^icb^j ~|~i\leq j\}$, which is well-known  not to be regular \cite{hopcroft2001introduction}, it is immediate that Lexicographic Contraction cannot be encoded by a PDL progam.
\end{proof}

\begin{refprop}{\ref{prop:Faith}}
Let $\mathfrak{C}$ be a class of dynamic operators $\star$ satisfying $\mathtt{(Faith)}$ and $\mathfrak{M}\subseteq \mathit{Mod}(\mathcal{L}_\leq)$ be a class of preference models. The following axiom schema is valid in $\langle \mathfrak{M}, \mathfrak{C}\rangle$, for any propositional formula $\varphi \in \mathcal{L}_0$.
\vspace{-0.2cm}
$$
\begin{array}{lcl}
E\varphi  & \rightarrow  & \mu\varphi \leftrightarrow [\star\varphi]\mu\top\\
\end{array}
$$
\end{refprop}
\begin{proof}
Let $D = \langle M, \star\rangle$ be a dynamic model, s.t. $\star: \mathit{Mod}(\mathcal{L}_\leq)\times \mathcal{L}_0 \rightarrow \mathit{Mod}(\mathcal{L}_\leq)$ is a dynamic operator satisfying \FAITH{} and $M=\langle W,\leq,v\rangle \in \mathfrak{M}$ is a preference model, $D-\langle M,\star\rangle$, and let $\varphi \in \mathcal{L}_0$ be a propositional formula s.t. $\llbracket \varphi\rrbracket \neq \emptyset$. Let's call $\star(M,\varphi) = M_{\star\varphi}= \langle W, \leq_{\star\varphi},v\rangle$ and $D'=\langle M_{\star \varphi},\star\rangle$.

Suppose $D\vDash E\varphi$, then $\llbracket \varphi\rrbracket \neq \emptyset$. As $M$ is well-founded, $Min_\leq \llbracket \varphi \rrbracket \neq \emptyset$ i.e. there is some $w \in W$ s.t. $D,w\vDash \mu\varphi$. Take one such $w$, since $\star$ satisfies \FAITH{}, then $w\in Min_{\leq_{\star \varphi}} W$. Then, $D', w\vDash \mu \top$, i.e. $D,w\vDash [\star \varphi]\mu \top$. 
\end{proof}

\begin{refprop}{\ref{prop:charFaith}}
Let $\mathfrak{D}=\langle \mathfrak{M}, \star\rangle$ be a class of dynamic models. The axiom schema in Proposition~\ref{prop:Faith} is valid in $\mathfrak{D}$, for any propositional formula $\varphi \in \mathcal{L}_0$ iff 
for each $M = \langle W, \leq, v\rangle \in \mathfrak{M}$, 
%
%
s.t. $\star(M, \varphi) = \langle W, \leq_{\star \varphi}, v\rangle$, if $\llbracket \varphi\rrbracket\neq \emptyset$, 
then  $Min_\leq \llbracket \varphi \rrbracket = Min_{\leq_{\star \varphi}} W$,
i.e. $\star$ satisfies \FAITH{} in $\mathfrak{M}$.
\end{refprop}
\begin{proof}
Let $\mathfrak{D}=\langle \mathfrak{M}, \star\rangle$ be a class of dynamic models for which the axiom schema in Proposition~\ref{prop:Faith} is valid. Let $D =\langle M, \star\rangle \in\mathfrak{D}$ be a dynamic model with $M = \langle W, \leq, v\rangle$, $\varphi \in \mathcal{L}_0$ be a propositional formula s.t. $\llbracket \varphi \rrbracket \neq \emptyset$, and $D' = \langle M_{\star\varphi}, \star\rangle$ with $\star(M, \varphi) = M_{\star \varphi}= \langle W, \leq_{\star \varphi}, v\rangle$.

$i)~ Min_{\leq} \llbracket \varphi \rrbracket \subseteq Min_{\leq_{\star \varphi}} W$:

Notice that, since $\mathfrak{M}\subseteq \mathit{Mod}(\mathcal{L}_\leq)$, i.e. it is a well-founded preference model, and $\llbracket \varphi \rrbracket\neq \emptyset$, by well-foundedness $Min_\leq \llbracket \varphi \rrbracket\neq \emptyset$. As such, take $w\in {Min_{\leq} \llbracket \varphi \rrbracket}$, then  $D,w\vDash \varphi$, which implies $D,w\vDash E\varphi$. Also, since $w\in {Min_{\leq} \llbracket \varphi \rrbracket}$, by Proposition~\ref{def:mu}, it holds that $D,w\vDash \mu \varphi$. Hence, $M,w\vDash E\varphi \wedge \mu \varphi$. Since the axiom schema is valid in $\mathfrak{D}$, it must hold that $D,w\vDash [\star\varphi]\mu\top$, i.e. $D',w\vDash \mu \top$. By Proposition~\ref{def:mu}, we conclude that $w\in Min_{\leq_{\star \varphi}} \llbracket \top\rrbracket = Min_{\leq_{\star \varphi}} W$. As such, $Min_\leq \llbracket \varphi \rrbracket \subseteq Min_{\leq_{\star \varphi}} W$.

$ii)~ Min_{\leq} \llbracket \varphi \rrbracket \supseteq Min_{\leq_{\star \varphi}} W$:  

Notice that, since $\mathfrak{M}\subseteq \mathit{Mod}(\mathcal{L}_\leq)$, i.e. it is a well-founded preference model, and $W\neq \emptyset$, by well-foundedness $Min_{\leq_{\star \varphi}} W\neq \emptyset$. As such, take $w\in Min_{\leq_{\star \varphi}} W$. By Proposition~\ref{def:mu}, it holds that $D',w\vDash \mu \top$, i.e. $D,w\vDash [\star\varphi]\mu\top$. Since the axiom schema is valid in $\mathfrak{D}$, it must hold that $D,w\vDash \mu\varphi$. By Proposition~\ref{def:mu}, $w\in Min_{\leq} \llbracket \varphi\rrbracket$. As such, $Min_{\leq_{\star \varphi}} W\subseteq Min_\leq \llbracket \varphi \rrbracket$.
\end{proof}

\begin{refprop}{\ref{prop:CR1}}\sloppy
Let $\mathfrak{C}$ be a class of dynamic operators $\star$ satisfying \DP{1} and $\mathfrak{M}\subseteq \mathit{Mod}(\mathcal{L}_\leq)$ be a class of preference models. The following axiom schemata is valid in $\langle \mathfrak{M}, \mathfrak{C}\rangle$ for any $\varphi \in \mathcal{L}_0$ and $\xi \in \mathcal{L}_\leq(\star)$.
\vspace{-0.2cm}
$$
\begin{array}{lcl}
{}[\star\varphi][\leq]\xi &\rightarrow& (\varphi \rightarrow [\leq](\varphi \rightarrow [\star\varphi] \xi))\\
{}[\star\varphi][<]\xi &\rightarrow& (\varphi \rightarrow [<](\varphi \rightarrow [\star\varphi] \xi))\\
{}[\leq][\star\varphi]\xi &\rightarrow& (\varphi \rightarrow [\star\varphi][\leq](\varphi \rightarrow \xi))\\
{}[<][\star\varphi]\xi &\rightarrow& (\varphi \rightarrow [\star\varphi][<](\varphi \rightarrow \xi))\\
\end{array}$$
\end{refprop}
\begin{proof}\sloppy
We will only show the case for the axioms regarding the modality $[\leq]$ (axioms 1 and 3), since for the axioms involving modality $[<]$, it suffices to observe that \DP{1} implies that for any $w,w'\in \llbracket \varphi\rrbracket$ it holds that $w<w'$ iff $w<_{\star \varphi} w'$.

Let $M = \langle W, \leq, v\rangle \in \mathfrak{M}$ be a preference model, $\star \in \mathfrak{C}$ be a dynamic operator satisfying \DP{1}, and $\varphi \in \mathcal{L}_0$ be a propositional formula. For the sake of presentation, let's call $D = \langle M, \star \rangle$ and $D' = \langle M_{\star \varphi}, \star \rangle$, with $M_{\star \varphi} = \star(M,\varphi) = \langle W, \leq_{\star \varphi}, v\rangle$.

(i) Take $w\in W$ and $\xi \in \mathcal{L}_\leq(\star)$ s.t. $D,w\vDash [\star\varphi][\leq]\xi$. Clearly, if $D,w\not\vDash \varphi$, it holds that $D,w\vDash \varphi \rightarrow [\leq](\varphi \rightarrow  [\star\varphi] \xi)$, so we only need to consider the case in which $D,w \vDash \varphi$. Since $D,w\vDash [\star\varphi][\leq]\xi$ , by Definition~\ref{def:semantics}, $D',w\vDash [\leq]\xi$, i.e. for any $w'\in W$ s.t. $w'\leq_{\star\varphi}w$, it holds that $D',w'\vDash \xi$. Then, it holds that $D,w'\vDash [\star\varphi]\xi$. Notice that, by Proposition~\ref{prop:BasicAxiom}, for any $w'\in W$ it holds that $D',w'\vDash \varphi$  iff $D,w'\vDash \varphi$, since $\varphi$ is a propositional formula. As such, take $w'\in W$ s.t. $w' \leq w$. If $D,w'\not\vDash \varphi$, then $D,w' \vDash \varphi \rightarrow [\star\varphi] \xi$. Otherwise, if $D,w'\vDash \varphi$, then since $\star$ satisfies \DP{1}, it holds that $w' \leq_{\star \varphi} w$ and $D',w'\vDash \xi$, thus $D,w'\vDash \varphi \rightarrow [\star \varphi]\xi$. As such, by Definition~\ref{def:semantics}, $D,w\vDash [\leq](\varphi \rightarrow  [\star\varphi] \xi)$. Since $D,w\vDash \varphi$, it holds that $D,w\vDash \varphi \rightarrow [\leq](\varphi \rightarrow  [\star\varphi] \xi)$.

(ii) Take $w\in W$ and $\xi \in \mathcal{L}_\leq(\star)$ s.t. $D,w\vDash [\leq][\star\varphi]\xi$. Again, if $D,w\not\vDash \varphi$, it holds immediately that $D,w \vDash (\varphi \rightarrow [\star\varphi][\leq](\varphi \rightarrow \xi))$, so let's assume $D,w\vDash \varphi$. By $D,w\vDash [\leq][\star\varphi]\xi$, we conclude that for all $w'\in W$ s.t. $w'\leq w$ it holds that $D,w'\vDash [\star\varphi]\xi$, i.e $D',w'\vDash \xi$. Since $\star$ satisfies \DP{1}, for any world $w' \in W$ s.t. $D,w'\vDash \varphi$ , it holds that $w'\leq w$ iff $w' \leq_{\star \varphi} w$. Take $w'\in W$ s.t. $w'\leq_{\star \varphi} w$. Clearly, if $D',w'\not\vDash \varphi$ then $D',w'\vDash (\varphi \rightarrow \xi)$. Otherwise, if $D',w'\vDash \varphi$, then $w'\leq w$ and, thus, $D',w'\vDash \xi$, i.e. $D',w'\vDash \varphi \rightarrow \xi$. A such, $D',w\vDash [\leq](\varphi \rightarrow \xi)$ and we conclude, by Definition~\ref{def:semantics}, that ${D,w\vDash [\star \varphi][\leq](\varphi \rightarrow \xi)}$. Since $D,w\vDash \varphi$, then $D,w\vDash \varphi \rightarrow [\star \varphi][\leq](\varphi \rightarrow \xi)$.
\end{proof}

\begin{reffact}{\ref{fact:cr1}}\sloppy
There are two dynamic operators $\star,\ast:\mathit{Mod}(\mathcal{L}_\leq)\times \mathcal{L}_0 \rightarrow \mathit{Mod}(\mathcal{L}_\leq)$ s.t. for any class of preference models $\mathfrak{M}$, s.t. both $\star$ and $\ast$ are closed over $\mathfrak{M}$, and formula $\xi \in \mathcal{L}_\leq(\star)$, $\xi$ is satisfiable in $\langle \mathfrak{M}, \star\rangle$ iff $\xi$ is satisfiable in $\langle \mathfrak{M}, \ast\rangle$, but $\star$ does not satisfies \DP{1} while $\ast$ does.
\end{reffact}
\begin{proof}
Let $W =\{w_1,w_2\}$, consider the preference models $M_1 = \langle W, \leq_1, v\rangle$ and $M_2 = \langle W, \leq_2, v\rangle$ with $$\leq_1 =\{\langle w1,w1 \rangle, \langle w2,w2 \rangle\}$$ and $$\leq_2 =\{\langle w1,w1 \rangle, \langle w1,w2 \rangle, \langle w2,w1 \rangle, \langle w2,w2 \rangle\}$$ and any propositional valuation s.t. for any $p\in P$, $w_1 \in v(p)$ iff $w_2 \in v(p)$. 

Clearly, $M_1$ is bissimilar to $M_2$ \cite{blackburn2006handbook}, and for any $w\in W$ and $\xi \in \mathcal{L}_\leq$, $M_1,w\vDash \xi$ iff $M_2,w\vDash \xi$. Let's define the operations $\star,\ast:\mathit{Mod}(\mathcal{L}_\leq)\times \mathcal{L}_0 \rightarrow \mathit{Mod}(\mathcal{L}_\leq)$ as
$$
\begin{array}{ll}
\ast(M,\varphi) =& M\\
\star(M,\varphi) =&\begin{cases}
M & \mbox{if }M \neq M_1 \mbox{ or } \varphi \neq \top\\
M_2 & \mbox{if }M = M_1 \mbox{ and } \varphi = \top\\
\end{cases}\end{array}$$

Notice that, since $M_1$ and $M_2$ are bissimilar, for any $M\in \mathit{Mod}(\mathcal{L}_\leq)$ and $\varphi \in \mathcal{L}_0$, $\ast(M, \varphi)$ is bissimilar to $\star(M,\varphi)$ and, thus, modally equivalent \cite{blackburn2006handbook}. As such, for any class of preference models $\mathfrak{M} \subseteq \mathit{Mod}(\mathcal{L}_\leq)$ s.t. $\star$ is closed over $\mathfrak{M}$\footnote{$\star$ is clearly closed for any class $\mathfrak{M}$ s.t. either $M_1 \not \in \mathfrak{M}$ or $M_1,M_2 \in \mathfrak{M}$.}, it holds that $\xi \in \mathcal{L}_\leq(\star)$ is satisfiable in $\langle \mathfrak{M}, \star\rangle$ iff it is satisfiable in $\langle \mathfrak{M}, \ast\rangle$. Also, it is clear that $\ast$ satisfies \DP{1} and $\star$ does not, since for $w_2 \leq_2 w_1$ but $w_2 \not\leq_1 w_1$. 
\end{proof}

\textcolor{blue}{\begin{reffact}{\ref{fact:dp1syn}}
Let $\mathfrak{D} = \langle \mathfrak{M}, \star\rangle$ be a class of dynamic models satisfying the axiom schemata in Proposition~\ref{prop:CR1}. For any propositional formulas $\varphi, \psi \in \mathcal{L}_0$ s.t. ${\mathfrak{D}\vDash \varphi \rightarrow \psi}$ and any dynamic preference formula $\xi \in \mathcal{L}_\leq(\star)$, it holds that
$$\mathfrak{D}\vDash [\star \psi]B(\xi~|~\varphi) \leftrightarrow B([\star \psi] \xi ~|~\varphi)$$
\end{reffact}}
\textcolor{blue}{
\begin{proof}
Since $\mathfrak{D}$ is a class of dynamic models satisfying the axiom schemata in Proposition~\ref{prop:CR1}, it holds that
$$
\begin{array}{lll}
\mathfrak{D}\vDash[\star \psi]B(\xi~|~\varphi) &\equiv_{def} & [\star \psi]A(\mu \varphi \rightarrow \xi)\\
                             &\equiv_{def} & [\star \psi]A((\varphi\wedge \neg\langle < \rangle\varphi) \rightarrow \xi)\\
                             &\leftrightarrow_{Prop~\ref{prop:BasicAxiom}} & A([\star \psi]\varphi \wedge [\star \psi][<]\neg \varphi \rightarrow [\star \psi]\xi)\\
                             &\leftrightarrow_{Prop~\ref{prop:CR1}} & A((\varphi \wedge (\psi \rightarrow [<]( \psi \rightarrow [\star \psi]\neg \varphi))) \rightarrow [\star \psi]\xi)\\
                             &\leftrightarrow_{Prop~\ref{prop:BasicAxiom}} & A((\varphi \wedge (\psi \rightarrow [<]( \psi \rightarrow\neg \varphi))) \rightarrow [\star \psi]\xi)\\                             
                             &\leftrightarrow_{\mathfrak{D}\vDash \varphi \rightarrow \psi} & A((\varphi \wedge [<]\neg \varphi) \rightarrow [\star \psi]\xi)\\
                             &\equiv_{def} & A((\varphi \wedge \neg \langle <\rangle \varphi) \rightarrow [\star \psi]\xi)\\          
                             &\equiv_{def} & A(\mu\varphi \rightarrow [\star \psi]\xi)\\          
                             &\equiv_{def} & B([\star \psi]\xi~|~\varphi)\\          
\end{array}$$
\end{proof}
} 

\begin{refprop}{\ref{prop:dp1plus}}\sloppy
Let $\mathfrak{D} = \langle \mathfrak{M}, \star\rangle$ be a class of dynamic models. The axiomatisation presented in Proposition~\ref{prop:CR1} is valid in $\mathfrak{D}$ iff $\star$ is $\mathfrak{M}$-DP1-compliant.
\end{refprop}
\begin{proof}
Let $D = \langle M, \star\rangle \in \mathfrak{D}$ s.t. $M = \langle W, \leq, v\rangle$ is a preference model and $\star$ be a dynamic operator. For any $\varphi \in \mathcal{L}_0$, let's call $\star(M,\varphi)=M_{\star \varphi} = \langle W, \leq_{\star \varphi}, v\rangle$ and $D' = \langle M_{\star \varphi}, \star\rangle$.

$\Rightarrow$:

(i) Take $w,w'\in \llbracket \varphi \rrbracket$ s.t. $w\leq_{\star \varphi} w'$ ($w<_{\star \varphi} w'$) and $D,w\vDash [\star \varphi]\xi$ for some $\xi \in \mathcal{L}_\leq(\star)$,  then $D,w\vDash \varphi \wedge [\star \varphi](\varphi \wedge \xi)$ and $D,w'\vDash \varphi \wedge \langle \leq \rangle (\varphi \wedge [\star \varphi](\varphi \wedge \xi))$ (or $D,w'\vDash \varphi \wedge \langle < \rangle (\varphi \wedge [\star \varphi](\varphi \wedge \xi)$, if $w<_{\star \varphi} w'$). Since the axiomatisation presented in Proposition~\ref{prop:CR1} is valid in $\mathfrak{D}$, it holds by contraposition of the third (fourth) axiom that $D,w'\vDash \langle \leq \rangle [\star \varphi] (\varphi \wedge\xi)$ ($D,w'\vDash \langle < \rangle [\star \varphi] (\varphi \wedge\xi)$). As such, there is a $w''\in \llbracket \varphi\rrbracket$ s.t. $w''\leq w'$ ($w''< w'$) and $D,w''\vDash [\star \varphi] \xi$. In other words, $\star$ satisfies \DP{1a} in $\mathfrak{M}$.

(ii) Take $w,w'\in \llbracket \varphi \rrbracket$ s.t. $w\leq w'$ ($w< w'$) and $D,w\vDash [\star \varphi]\xi$ for some $\xi \in \mathcal{L}_\leq(\star)$. Then $D,w\vDash [\star \varphi](\varphi \wedge \xi)$ and ${D,w'\vDash \varphi \wedge \langle \leq \rangle [\star \varphi](\varphi \wedge \xi)}$ (similarly, ${D,w'\vDash \varphi \wedge \langle < \rangle [\star \varphi](\varphi \wedge \xi)}$ if $w<w'$). Since the axiomatisation in Proposition~\ref{prop:CR1} is valid in $\mathfrak{D}$, by contraposition of the first (second) axiom, then ${D,w'\vDash [\star \varphi] \langle \leq \rangle (\varphi \wedge\xi)}$ (${D,w'\vDash [\star \varphi] \langle < \rangle (\varphi \wedge\xi)}$). In other words, there is some $w'' \in \llbracket \varphi \rrbracket$, s.t. $D,w''\vDash [\star \varphi] \xi$ and $w''\leq_{\star \varphi} w'$ ($w''<_{\star \varphi} w'$). Thus, $\star$ satisfies \DP{2a} in $\mathfrak{M}$.

$\Leftarrow$:

As before, we will omit the proof for the axioms regarding the modality $[<]$ since they are similar to that of $[\leq]$.

(i) Take $w\in W$ s.t. ${D,w\vDash \varphi \wedge [\star \varphi] \langle \leq \rangle (\varphi \wedge \xi)}$, for some $\varphi \in \mathcal{L}_0$ and $\xi\in \mathcal{L}_\leq(\star)$. Then, $w \in \llbracket \varphi \rrbracket$ and there is some world $w'\in \llbracket \varphi\rrbracket$ s.t. $D,w'\vDash [\star \varphi]\xi$ and $w' \leq_{\star \varphi} w$. By \DP{1a}, there is some world $w''\in \llbracket \varphi \rrbracket$ s.t. $D,w''\vDash [\star\varphi] \xi$ and $w''\leq w$. Then $D,w\vDash \langle \leq \rangle [\star\varphi]\xi$. As such, we conclude that $$D\vDash (\varphi \wedge [\star \varphi] \langle \leq \rangle (\varphi \wedge \xi)) \rightarrow  \langle \leq \rangle [\star\varphi]\xi.$$ By the contrapositive, $$D\vDash [\leq][\star\varphi]\xi \rightarrow(\varphi \rightarrow [\star\varphi][\leq](\varphi \rightarrow \xi)).$$ Thus, $$\mathfrak{D}\vDash [\leq][\star\varphi]\xi \rightarrow (\varphi \rightarrow [\star\varphi][\leq](\varphi \rightarrow \xi)).$$

(ii) Take $w\in W$ s.t. ${D,w\vDash \varphi \wedge \langle \leq \rangle (\varphi \wedge [\star \varphi](\varphi \wedge \xi))}$, for some $\varphi \in \mathcal{L}_0$ and $\xi\in \mathcal{L}_\leq(\star)$. Then $w \in \llbracket \varphi \rrbracket$ and there is some world $w'\in \llbracket \varphi\rrbracket$ s.t. $D,w'\vDash [\star \varphi]\xi$ and $w' \leq  w$. By \DP{1b}, there is some world $w''\in \llbracket \varphi \rrbracket$ s.t. $D,w''\vDash [\star\varphi] \xi$ and $w''\leq_{\star\varphi} w$. As such, ${D,w\vDash [\star \varphi]\langle \leq \rangle \xi}$ and, as such, $${D\vDash (\varphi \wedge \langle \leq \rangle (\varphi \wedge [\star \varphi](\varphi \wedge \xi)))\rightarrow [\star \varphi]\langle \leq \rangle \xi}.$$ By the contrapositive, $${D\vDash [\star\varphi][\leq]\xi \rightarrow (\varphi \rightarrow [\leq](\varphi \rightarrow [\star\varphi] \xi))}$$ and thus, $${\mathfrak{D}\vDash [\star\varphi][\leq]\xi \rightarrow (\varphi \rightarrow [\leq](\varphi \rightarrow [\star\varphi] \xi))}.$$
\end{proof}

\textcolor{blue}{
\begin{reffact}{\ref{cor:CR1}}
Let $\mathfrak{M}$ be a class of preference models s.t. for any $M\in\mathfrak{M}$ and any possible world $w$ in $M$ there is a characteristic formula $\xi_w\in \mathcal{L}_\leq$, s.t. $M,w'\vDash \xi_w$ iff $w'=w$ and let $\star:Mod(\mathcal{L}_\leq)\times \mathcal{L}_0 \rightarrow Mod(\mathcal{L}_\leq)$ be a dynamic operator closed over $\mathfrak{M}$. It holds that $\star$ is $\mathfrak{M}$-DP1-compliant iff for any  $M\in \mathfrak{M}$, any propositional formula $\varphi$ and worlds $w,w'\in \llbracket\varphi\rrbracket$ it holds that $w\leq w'$ iff $w \leq_{\star \varphi} w'$.
\end{reffact}}
\begin{proof}
\textcolor{blue}{Notice that the implication that satisfaction of the \DP{1} condition implies DP1-compliance is trivial and holds for any class of models, so we will focus on implication that for adequate classes of models $\mathfrak{M}$, which includes the class of all Grove spheres (or concrete models in the terminology of \cite{souza:dali19}), DP1-compliance implies satisfaction of the \DP{1} condition.}

\textcolor{blue}{Take $\mathfrak{M}$ a class of preference models with characteristic formulas and $\star$ be a $\mathfrak{M}$-DP1-compliant dynamic operator closed over $\mathfrak{M}$. Let $M = \langle W, \leq, v\rangle \in \mathfrak{M}$ be a preference model and $w,w'\in W$ be possible worlds s.t. $w,w'\in \llbracket \varphi \rrbracket$. Let's call $\star(M,\varphi) = M_{\star\varphi}= \langle W, \leq_{\star \varphi}, v\rangle$. We need to show that   $w\leq w'$ iff $w \leq_{\star \varphi} w'$.}

\textcolor{blue}{Suppose $w\leq w'$. As $\star$ is closed over $\mathfrak{M}$, then $M_{\star \varphi}\in \mathfrak{M}$, thus there is $\xi_{w}\in\mathcal{L}_\leq$ s.t. $M_{\star\varphi},w''\vDash \xi_{w}$ iff $w''=w$. Clearly,  $M_{\star\varphi},w\vDash \xi_{w}$. By \DP{1b} we conclude that there is $w''\in W$ s.t. $M,w''\vDash [\star \varphi]\xi_w$ and $w''\leq_{\star \varphi} w'$, but this implies  $M_{\star \varphi},w''\vDash \xi_w$ and $w''\leq_{\star \varphi} w'$. Since $\xi_w$ is the characteristic formula of $w$ in $M_{\star \varphi}$, we conclude that $w\leq_{\star \varphi} w'$.}

\textcolor{blue}{Now, suppose $w\leq_{\star\varphi} w'$. As $\star$ is closed over $\mathfrak{M}$, then $M_{\star \varphi}\in \mathfrak{M}$, thus there is $\xi_{w}\in\mathcal{L}_\leq$ s.t. $M_{\star\varphi},w''\vDash \xi_{w}$ iff $w''=w$. Clearly,  $M_{\star\varphi},w\vDash \xi_{w}$. By \DP{1a} we conclude that there is $w''\in W$ s.t. $M,w''\vDash [\star \varphi]\xi_w$ and $w''\leq w'$, but this implies  $M_{\star \varphi},w''\vDash \xi_w$ and $w''\leq w'$. Since $\xi_w$ is the characteristic formula of $w$ in $M_{\star \varphi}$, we conclude that $w\leq w'$.}
\end{proof}

%

\begin{refprop}{\ref{prop:CR3}}
Let $\mathfrak{D} = \langle \mathfrak{M}, \star\rangle$ be a class of dynamic models. The following axiom schema is valid in $\mathfrak{D}$  for any $\varphi \in \mathcal{L}_0$ and $\xi \in \mathcal{L}_\leq(\star)$ iff $\star$ is $\mathfrak{M}$-DP3-compliant.
\vspace{-0.2cm}
$$
\begin{array}{lcl}
{}[\star \varphi][<](\varphi \rightarrow \xi) & \rightarrow &\neg \varphi \rightarrow [<] [\star \varphi] (\varphi \rightarrow \xi)
\end{array}$$
\end{refprop}
\begin{proof}
Let $D = \langle M, \star\rangle \in \mathfrak{D}$ s.t. $M = \langle W, \leq, v\rangle$ is a preference model and $\star$ be a dynamic operator. For any $\varphi \in \mathcal{L}_0$, let's call $\star(M,\varphi)=M_{\star \varphi} = \langle W, \leq_{\star \varphi}, v\rangle$ and $D' = \langle M_{\star \varphi}, \star\rangle$.

$\Rightarrow$:

Take $w,w' \in W$ s.t. $w\in \llbracket \varphi \rrbracket$, $w' \not\in \llbracket \varphi \rrbracket$, and $w < w'$, and $\xi \in \mathcal{L}_\leq(\star)$ s.t. $D',w \vDash \xi$. Then, $D,w\vDash [\star \varphi] (\varphi \wedge \xi)$, therefore, $$D,w' \vDash  \neg \varphi \wedge \langle < \rangle [\star \varphi] (\varphi \wedge \xi).$$ By hypothesis, it holds $\mathfrak{D}\vDash [\star \varphi][<](\varphi \rightarrow \xi)  \rightarrow \neg \varphi \rightarrow [<] [\star \varphi] (\varphi \rightarrow \xi)$. Then, by contraposition, $D,w\vDash [\star \varphi]\langle < \rangle (\varphi \wedge \xi)$, i.e. $D',w'\vDash \langle < \rangle (\varphi \wedge \xi)$. As such, there is some $w''\in W$ s.t. $w'' \in \llbracket \varphi \rrbracket$ and $D',w''\vDash \xi$, hence \DP{3a} holds. As such, we conclude that $\star$ is $\mathfrak{M}$-DP3-compliant.

$\Leftarrow$:

Take $w\in W$ s.t. $D,w \vDash \neg \varphi \wedge \langle < \rangle [\star \varphi ](\varphi \wedge \xi)$ for some $\xi \in\mathcal{L}_\leq(\star)$. Thus, $w \not\in \llbracket \varphi \rrbracket$ and there is some $w'\in W$ s.t. $w'\in \llbracket \varphi \rrbracket$, $D',w'\vDash \xi$. Since $\star$ is $\mathfrak{M}$-DP3-compliant, it holds that there is $w''\in W$ s.t. $w'' <_{\star \varphi} w'$ and $D',w''\vDash \xi$. As such, $D,w\vDash [\star \varphi]\langle < \rangle (\varphi \wedge \xi)$. Hence, we conclude that $$D\vDash  (\neg \varphi \wedge \langle < \rangle [\star \varphi ](\varphi \wedge \xi))\rightarrow [\star \varphi]\langle < \rangle (\varphi \wedge \xi).$$
By contraposition, $$D\vDash ([\star \varphi][<](\varphi \rightarrow \xi))  \rightarrow (\neg \varphi \rightarrow [<] [\star \varphi] (\varphi \rightarrow \xi)).$$
Thus, $$\mathfrak{D}\vDash ([\star \varphi][<](\varphi \rightarrow \xi))  \rightarrow (\neg \varphi \rightarrow [<] [\star \varphi] (\varphi \rightarrow \xi)).$$
\end{proof}

\textcolor{blue}{\begin{reffact}{\ref{fact:dp3syn}}
Let $\mathfrak{M}\subseteq \mathit{Mod}(\mathcal{L}_\leq)$ be a class of preference models, $\star: \mathit{Mod}(\mathcal{L}_\leq)\times \mathcal{L}_0 \rightarrow \mathit{Mod}(\mathcal{L}_\leq)$ be a $\mathfrak{M}$-DP3-compliant dynamic operator, and $\mathfrak{D} = \langle \mathfrak{M}, \star\rangle$. For any propositional formula $\varphi \in \mathcal{L}_0$ and dynamic preference formula $\xi \in \mathcal{L}_\leq(\star)$, it holds that
$$\mathfrak{D}\vDash B(\varphi~|~[\star \varphi]\xi) \rightarrow [\star \varphi]B(\varphi ~|~\xi)$$
\end{reffact}}

\begin{proof}
\textcolor{blue}{Let $D = \langle M, \star\rangle \in \mathfrak{D}$ s.t. $M = \langle W, \leq, v\rangle$ is a preference model and $\star$ be a dynamic operator and $D\vDash B(\varphi~|~[\star \varphi]\xi)$. For any $\varphi \in \mathcal{L}_0$, let's call $\star(M,\varphi)=M_{\star \varphi} = \langle W, \leq_{\star \varphi}, v\rangle$ and $D' = \langle M_{\star \varphi}, \star\rangle$. We have to show that $Min_{\leq_{\star\varphi}} \llbracket \xi \rrbracket_{D'} \subseteq \llbracket \varphi \rrbracket_{D'}$.}

\textcolor{blue}{
Take $w\in Min_{\leq_{\star \varphi}} \llbracket \xi\rrbracket_{D'}$ and suppose ${D',w\not\vDash \varphi}$, by definition ${D',w\vDash [<]\neg \xi}$, thus by Proposition~\ref{prop:BasicAxiom}, $D,w\vDash [\star \varphi][<]\neg\xi$. Since $\star$ is $\mathfrak{M}$-DP3-compliant, by Proposition~\ref{prop:CR3}, $D,w\vDash [<][\star \varphi](\xi\rightarrow \neg \varphi)$. As $D\vDash B(\varphi~|~[\star \varphi]\xi)$, then $Min_{\leq} \llbracket [\star \varphi]\xi \rrbracket_{D} \subseteq \llbracket \varphi \rrbracket_{D}$ and thus either (i) $w\in \llbracket [\star \varphi]\xi \rrbracket_{D}$ which contradicts $D',w\not\vDash \varphi$, or (ii) $Min_{\leq} \llbracket [\star \varphi]\xi \rrbracket_{D} \not\subseteq \llbracket \varphi \rrbracket_{D}$, which contradicts $D\vDash B(\varphi~|~[\star \varphi]\xi)$. Thus it must hold that $D',w\vDash \varphi$, i.e. $Min_{\leq_{\star \varphi}} \llbracket \xi\rrbracket_{D'} \subseteq \llbracket \varphi\rrbracket_{D'}$ and, thus, $D\vDash [\star \varphi]B(\varphi ~|~\xi)$.}
\end{proof}

%

\begin{refprop}{\ref{prop:Rec}}
Let $\mathfrak{D} = \langle \mathfrak{M}, \star\rangle$ be a class of dynamic models. The following axiom schemata  is valid in $\mathfrak{D}$  for any $\varphi \in \mathcal{L}_0$ and $\xi \in \mathcal{L}_\leq(\star)$ iff $\star$ is $\mathfrak{M}$-Rec-compliant.
\vspace{-0.2cm}
$$
\begin{array}{lcl}
{}[\star\varphi][<]\xi &\rightarrow& (\neg \varphi \rightarrow A(\varphi \rightarrow [\star\varphi] \xi))\\
{}\varphi &\rightarrow &([\star\varphi][\leq]\varphi)\\

\end{array}$$
\end{refprop}

\begin{proof}
Let $D = \langle M, \star\rangle \in \mathfrak{D}$ s.t. $M = \langle W, \leq, v\rangle$ is a preference model and $\star$ be a dynamic operator. For any $\varphi \in \mathcal{L}_0$, let's call $\star(M,\varphi)=M_{\star \varphi} = \langle W, \leq_{\star \varphi}, v\rangle$ and $D' = \langle M_{\star \varphi}, \star\rangle$.

$\Rightarrow$:

Take $w,w'\in W$ s.t. $w \in \llbracket\varphi\rrbracket$ and $w'\not\in \llbracket \varphi \rrbracket$ and $\xi \in \mathcal{L}(\star)$ s.t. ${D',w\vDash \xi}$, i.e. $D,w\vDash [\star \varphi]\xi$. 

Firstly, let's show that $w'\not\leq_{\star \varphi} w$. As $w \in \llbracket\varphi\rrbracket$, then $D,w\vDash \varphi$. Since the axiom schemata is valid in $\mathfrak{D}$, then $D,w\vDash [\star \varphi][\leq]\varphi$, i.e. for any $w''\in W$ if $w'' \leq_{\star \varphi} w$ then $D',w''\vDash \varphi$. Since $\varphi$ is propositional formula, then it must hold that for any $w''\in W$ if $w'' \leq_{\star \varphi} w$ then $D,w''\vDash \varphi$. As $w'\not\in \llbracket \varphi \rrbracket$, then $D,w'\not \vDash\varphi$ and, thus, $w'\not\leq_{\star \varphi} w$.
 
Now let's show that all information is preserved. As $\varphi$ is a propositional formula, by Proposition~\ref{prop:BasicAxiom}, it holds that  ${D,w\vDash \varphi \wedge [\star \varphi](\varphi \wedge \xi)}$. Since there is a world in $W$ that satisfies $\varphi \wedge [\star \varphi](\varphi \wedge \xi)$, by Definition~\ref{def:semantics}, we can conclude that $D,w'\vDash \neg \varphi \wedge E(\varphi \wedge [\star \varphi](\varphi \wedge \xi))$. Since the axiom schemata is valid in $\mathfrak{D}$ by hypothesis, it must hold that $D,w'\vDash [\star \varphi]\langle <\rangle (\varphi \wedge \xi)$, i.e. $D',w'\vDash \langle <\rangle (\varphi \wedge \xi)$. As such, there is some $w''\in W$ s.t. $D',w''\varphi \wedge \xi$ and $w''<_{\star\varphi}w'$. Hence, $w''\in \llbracket \varphi\rrbracket$, $D,w'\vDash [\star \varphi]\xi$ and $w''<_{\star \varphi} w'$.

In other words, $\star$ satisfies \REC{'}. Since it holds for any $D\in \mathfrak{D}$, $\star$ is $\mathfrak{M}$-Rec-compliant.

$\Leftarrow$:

(i) Take $w'\in W$ s.t. $D,w'\vDash \neg \varphi \wedge E(\varphi \wedge [\star \varphi]\xi)$ for some $\xi \in \mathcal{L}_\leq(\star)$. Then $w'\in \llbracket \varphi \rrbracket$ and there is some $w\in W$ s.t. $w\in \llbracket \varphi\rrbracket$ and $D,w\vDash [\star \varphi]\xi$, i.e. $D',w\vDash \xi$. By \REC{'}, there is some $w''\in W$ s.t. $w'' <_{\star \varphi} w'$ and $D',w''\vDash \xi$, as such $D',w'\vDash \langle < \rangle \xi$, i.e. $D,w'\vDash [\star \varphi] \langle < \rangle \xi$. Since it holds for any $w' \in W$, we can conclude that 
$$D\vDash (\neg \varphi \wedge E(\varphi \wedge [\star \varphi]\xi)) \rightarrow [\star \varphi] \langle < \rangle \xi.$$
By contraposition,
$$D\vDash [\star \varphi][<]\xi \rightarrow (\neg \varphi \rightarrow A(\varphi \rightarrow [\star \varphi]\xi)).$$
Since it holds for any $D\in \mathfrak{D}$,
$$\mathfrak{D}\vDash [\star \varphi][<]\xi \rightarrow (\neg \varphi \rightarrow A(\varphi \rightarrow [\star \varphi]\xi)).$$

(ii) Take $w,w'\in W$ s.t. $D,w\vDash \varphi$ and $w' \leq_{\star \varphi} w$. Clearly,  $D,w' \vDash \varphi$, otherwise by \REC{'} $w'\not<_{\star \varphi} w$. As such, $D',w\vDash [\leq] \varphi$, i.e. $D,w\vDash [\star\varphi][\leq]\varphi$. Since it holds for any $w\in W$ and $D\in \mathfrak{D}$, we conclude that $\mathfrak{D}\vDash \varphi \rightarrow [\star\varphi][\leq]\varphi$
\end{proof}


\begin{refprop}{\ref{prop:LC}}
Let $\mathfrak{D} = \langle \mathfrak{M}, \star\rangle$ be a class of dynamic models. The following axiom schemata is valid in $\mathfrak{D}$, for all $n \in \mathbb{N}$, $\varphi \in \mathcal{L}_0$, and $\xi \in\mathcal{L}_\leq(\star)$  if $\star$ is $\mathfrak{M}$-LC-compliant.
\vspace{-0.2cm}
$$
\begin{array}{lcll}

{} [\star \varphi][\leq] \xi &\rightarrow &\displaystyle\bigwedge_{i=1}^{n}\bigwedge_{j=i}^{n} \mu dg_{\varphi}(j) \rightarrow A (\mu dg_{\varphi}(i)\rightarrow  [\star \varphi]\xi) \wedge\\
&&\displaystyle\bigwedge_{i=1}^{n}\bigwedge_{j=i}^{n} \mu dg_{\neg \varphi}(j) \rightarrow A (\mu dg_{\neg \varphi}(i)\rightarrow  [\star \varphi]\xi) \wedge\\

&&\displaystyle\bigwedge_{i=1}^{n}\bigwedge_{j=i}^{n} \mu dg_\varphi(j) \rightarrow A (\mu dg_{\neg \varphi}(i)\rightarrow  [\star \varphi]\xi) \wedge\\
&&\displaystyle\bigwedge_{i=1}^{n} \bigwedge_{j=i}^{n} \mu dg_{\neg \varphi(j)} \rightarrow A (\mu dg_{\varphi}(i)\rightarrow  [\star \varphi]\xi)\\

{} [\star \varphi][<] \xi &\rightarrow &\displaystyle\bigwedge_{i=1}^{n}\bigwedge_{j=i+1}^{n} \mu dg_{\varphi}(j) \rightarrow A (\mu dg_{\varphi}(i)\rightarrow  [\star \varphi]\xi) \wedge\\
&&\displaystyle\bigwedge_{i=1}^{n}\bigwedge_{j=i+1}^{n} \mu dg_{\neg \varphi}(j) \rightarrow A (\mu dg_{\neg \varphi}(i)\rightarrow  [\star \varphi]\xi) \wedge\\
&&\displaystyle\bigwedge_{i=1}^{n}\bigwedge_{j=i+1}^{n} \mu dg_\varphi(j) \rightarrow A (\mu dg_{\neg \varphi}(i)\rightarrow  [\star \varphi]\xi) \wedge\\
&&\displaystyle\bigwedge_{i=1}^{n} \bigwedge_{j=i+1}^{n} \mu dg_{\neg \varphi}(j) \rightarrow A (\mu dg_{\varphi}(i)\rightarrow  [\star \varphi]\xi)\\

{} [\star \varphi][\leq] \xi &\leftarrow & (\mu dg_{\neg\varphi}(n) \vee \mu dg_\varphi(n))\wedge \displaystyle\bigwedge_{i=1}^{n} A (\mu dg_{\varphi}(i)\rightarrow  [\star \varphi]\xi) \wedge \\
&&\displaystyle\bigwedge_{i=1}^{n} A (\mu dg_{\neg \varphi}(i)\rightarrow  [\star \varphi]\xi)\\

{} [\star \varphi][<] \xi &\leftarrow & (\mu dg_{\neg\varphi}(n) \vee \mu dg_\varphi(n))\wedge\displaystyle\bigwedge_{i=1}^{n-1} A (\mu dg_{\varphi}(i)\rightarrow  [\star \varphi]\xi) \wedge\\
&&\displaystyle\bigwedge_{i=1}^{n-1} A (\mu dg_{\neg \varphi}(i)\rightarrow  [\star \varphi]\xi)\\

\end{array}$$
\end{refprop}
\begin{proof}
We will only show it holds for the schemata related to modality $[\leq]$, since the proof for the others is similar.

Let $\mathfrak{D}=\langle \mathfrak{M}, \star\rangle$ be a class of dynamic models and $\varphi \in \mathcal{L}_0$ be a propositional formula. Take $D\in\mathfrak{D}$ s.t. $D= \langle M, \star\rangle$ and $D'= \langle M_{\star \varphi},\star\rangle$, with $M = \langle W, \leq, v\rangle$ and $M_{\star \varphi} = \langle W, \leq_{\star \varphi}, v\rangle$.

(i) Let $w\in W$ be a possible world s.t. $D,w\vDash [\star \varphi][\leq]\xi$ for some $\xi \in \mathcal{L}_\leq(\star)$ and $D,w\vDash \mu dg_\chi (n)$ for $\chi \in \{\varphi, \neg \varphi\}$ and $n\in \mathbb{N}$ - notice that one such $n$ always exists since the models are well-founded. As $D,w\vDash \mu dg_\chi (n)$, by Lemma~\ref{lem:dg}, there is a maximal chain of $\chi$-worlds of size $n$, starting in a minimal $\chi$-world and ending in $w$. Since $D,w\vDash [\star \varphi][\leq]\xi$, for all $w'\in W$ s.t. $w' \leq_{\star \varphi} w$, then $D',w'\vDash \xi$. Since $\star$ is $\mathfrak{M}$-LC-compliant, by \LC{'}, $w' \leq_{\star \varphi} w$ iff there is a maximal chain of $\chi'$-worlds of size $n'$, with $\chi'\in \{\varphi, \neg \varphi\}$, starting in a minimal $\chi'$-world and ending in $w'$ and $n'\leq n$. From that we conclude that for any $i \leq n$ and $w''\in W$, if $D,w''\vDash \mu dg_\chi (i)$, then it must hold that $D',w''\vDash \xi$, i.e. $D,w''\vDash [\star \varphi] \xi$. Thus, $\begin{array}{llcll}
D\vDash&[\star \varphi][\leq] \xi &\rightarrow &\displaystyle\bigwedge_{i=1}^{n}\bigwedge_{j=i}^{n} \mu dg_{\varphi}(j) \rightarrow A (\mu dg_{\varphi}(i)\rightarrow  [\star \varphi]\xi) \wedge\\
&&&\displaystyle\bigwedge_{i=1}^{n}\bigwedge_{j=i}^{n} \mu dg_{\neg \varphi}(j) \rightarrow A (\mu dg_{\neg \varphi}(i)\rightarrow  [\star \varphi]\xi) \wedge\\
&&&\displaystyle\bigwedge_{i=1}^{n}\bigwedge_{j=i}^{n} \mu dg_\varphi(j) \rightarrow A (\mu dg_{\neg \varphi}(i)\rightarrow  [\star \varphi]\xi) \wedge\\
&&&\displaystyle\bigwedge_{i=1}^{n} \bigwedge_{j=i}^{n} \mu dg_{\neg \varphi(j)} \rightarrow A (\mu dg_{\varphi}(i)\rightarrow  [\star \varphi]\xi)\mbox{, for } n \in \mathbb{N}\\\end{array}$

(ii) Let $w\in W$ be a possible world s.t.  $D,w\vDash \mu dg_\chi (n)$ for $\chi \in \{\varphi, \neg \varphi\}$ and $n\in \mathbb{N}$ and for any $i\leq n$ it holds that $D,w\vDash A (\mu dg_{\chi'}(i)\rightarrow  [\star \varphi]\xi)$, for $\chi'\in \{\varphi, \neg \varphi\}$. As such, for any $w'\in W$, if $D,w'\vDash \mu dg_{\chi'}(i)$ then $D,w'\vDash [\star \varphi]\xi$, for any $i\leq n$ and $\chi'\in \{\varphi, \neg \varphi\}$. As $D,w\vDash \mu dg_\chi (n)$ there is a maximal chain of $\chi$-worlds of size $n$, starting in a minimal $\chi$-world and ending in $w$. Since $\star$ is $\mathfrak{M}$-LC-compliant, by \LC{'}, $w' \leq_{\star \varphi} w$ iff there is a maximal chain of $\chi'$-worlds of size $n'$, with $\chi'\in \{\varphi, \neg \varphi\}$, starting in a minimal $\chi'$-world and ending in $w'$ and $n'\leq n$. From that we conclude that for all $w'\in W$ s.t. $w'\leq_{\star \varphi} w$ it must hold that $D,w' \vDash [\star \varphi]\xi$. As such, $D,w \vDash [\leq][\star \varphi]\xi$. Thus,
$ \begin{array}{lllll}
D\vDash& [\star \varphi][\leq] \xi &\leftarrow & (\mu dg_{\neg\varphi}(n) \vee \mu dg_\varphi(n))\wedge \displaystyle\bigwedge_{i=1}^{n} A (\mu dg_{\varphi}(i)\rightarrow  [\star \varphi]\xi) \wedge\\
&&&\displaystyle\bigwedge_{i=1}^{n} A (\mu dg_{\neg \varphi}(i)\rightarrow  [\star \varphi]\xi)\\
\end{array}$
\end{proof}

\begin{refteo}{\ref{teo:correct}}
Let $\mathfrak{M}$ be a class of preference models, $\mathcal{C} = \{\mathfrak{C}_i ~|~i\in I\}$ be a family of classes of dynamic operators which are closed over $\mathfrak{M}$, and $\mathcal{A} = \{A_i ~|~ i\in I\}$ a family of sound axiom systems for $\mathcal{C}$, i.e. $Log(A_i)\subseteq Log(\langle \mathfrak{M}, \mathfrak{C}_i\rangle)$, both indexed by some set $I$. \textcolor{blue}{$$Log(\bigcup_{i\in I} A_i) \subseteq Log(\langle \mathfrak{M}, \bigcap_{i\in I}\mathfrak{C}_i)$$}
\end{refteo}
\begin{proof}
Let's call $\bigcap \mathcal{C} = \{\star: Mod(\mathcal{L}_\leq)\times \mathcal{L}_0 \rightarrow Mod(\mathcal{L}_\leq) ~|~ \star \in \mathfrak{C}_i \mbox{ for all }i\in I\}$. Notice that if $\bigcap\mathcal{C} = \emptyset$, then $Log(\{\langle M, \star\rangle ~|~M \in \mathfrak{M} \mbox{ and } \star \in \mathfrak{C}_i\mbox{ for all }i\in I\}) = Log(\langle \mathfrak{M}, \bigcap \mathcal{C}\rangle) = Log(\emptyset) =\mathcal{L}_\leq(\star)$ and, clearly, $Log(\bigcup_{i\in I} A_i)\subseteq \mathcal{L}_\leq(\star)$, so let's assume $\bigcap\mathcal{C} \neq \emptyset$.

Take $\star\in \bigcap\mathcal{C}$, then for all $i\in I$, it must be the case that $Log( \mathcal{A}_i) \subseteq Log(\langle\mathfrak{M},\star\rangle)$, since $Log(\bigcup_{i\in I} A_i) \subseteq Log(\langle\mathfrak{M},\mathfrak{C}_i\rangle) \subseteq Log(\langle\mathfrak{M},\star\rangle)$, by Proposition~\ref{prop:intersect}. Particularly, $\bigcup_{i\in I} A_i \subseteq Log(\langle \mathfrak{M},\star\rangle)$. 

It is easy to see that for any class of preference models $ \mathfrak{M}$ and dynamic operator $\star$, $Log(\langle \mathfrak{M},\star\rangle)$ is closed over \textit{modus ponens} and necessitation rules. Hence, by Definition~\ref{def:logA}, $Log(\bigcup_{i \in I} A_i) \subseteq Log(\langle \mathfrak{M},\star \rangle)$. Since, for any $\star\in \bigcap\mathcal{C}$, $Log(\bigcup_{i\in I} A_i) \subseteq Log(\langle \mathfrak{M},\star \rangle)$, then 
$$
Log(\bigcup_{i\in I} A_i) \subseteq Log(\{\langle M, \star\rangle ~|~M \in \mathfrak{M} \mbox{ and } \star \in \mathfrak{C}_i\mbox{ for all }i\in I\}).$$
\end{proof}